\documentclass[%
 reprint,
 amsmath,amssymb,
aps,
]{revtex4-2}

\usepackage{graphicx}
\usepackage{dcolumn}
\usepackage{bm}
\usepackage{xcolor}
\usepackage{multirow}
\usepackage[utf8]{inputenc}
\usepackage[T1]{fontenc}

\begin{document}

\preprint{APS/123-QED}

\title{Experimental studies of the three nucleon system dynamics in the proton induced deuteron breakup at 108 MeV}

\author{A.~Łobejko$^{1,2}$, E.~Stephan$^1$, I.~Ciepał$^3$, J.~Golak$^4$, N.~Kalantar--Nayestanaki$^5$, S.~Kistryn$^4$, B.~Kłos$^1$, A.~Kozela$^3$, P.~Kulessa$^3$, W.~Parol$^3$, R.~Skibiński$^4$, I.~Skwira-Chalot$^6$,  A.~Szadziński$^1$, A.~Wilczek$^1$, H.~Witała$^4$, B.~Włoch$^{3,7}$, J.~Zejma$^4$} 

\affiliation{%
$^1$Institute of Physics, University of Silesia, PL-41500 Chorz\'ow, Poland \\
$^2$Institute of Experimental Physics, Faculty of Mathematics, Physics and informatics, University of Gdańsk, PL-80308 Gdańsk, Poland \\
$^3$Institute of Nuclear Physics, PAS, PL-31342 Krak\'ow, Poland \\
$^4$Faculty of Physics, Astronomy and Applied Computer Science, Jagiellonian University, PL-30348 Krak\'ow, Poland \\
$^5$ESRIG, University of Groningen, 9747 AA, Groningen, The Netherlands  \\
$^6$Faculty of Physics, University of Warsaw, PL-02093 Warszawa, Poland \\
$^7$Université de Bordeaux, CNRS, LP2I Bordeaux, FR-33170 Gradignan, France
}

\date{\today}

\begin{abstract}
The differential cross sections for the $^2$H(p,pp)n reaction have been measured for 84 angular configurations of the outgoing protons in the range of polar angles from 13 to 33 degrees with a~proton beam of 108 MeV. Data have been collected in the Cyclotron Center Bronowice (CCB) at the Institute of Nuclear Physics PAS in Cracow, using the BINA detector setup. Analysis leading to determination of the breakup cross section values is described. Absolute normalization is obtained by normalization to the simultaneously measured $^2$H(p,d)p scattering events.Experimental results are compared to the state-of-the-art theoretical calculations. Global analysis shows significant influence of the Coulomb interaction and small effects of three nucleon force in the studied phase space region.
\end{abstract}

\maketitle

\section{Introduction}\label{Intro}

Scattering experiments in systems of three nucleons provide particularly rich and sensitive data for testing the state-of-the-art potentials of nuclear interactions, such as realistic potentials or potentials stemming from Chiral Effective Field Theory (ChEFT).  In view of the neglect of the nucleons' internal degrees of freedom, applying the realistic potentials to describe systems beyond two nucleons requires complementing them with many-body forces, in particular a~so-called 3-nucleon force (3NF). Several models of such an interaction have emerged \cite{Coon1981,Coon2001,Pudliner1997,Pieper2001}. Alternatively, the explicit $\Delta$-isobar has been introduced in the coupled channel approach, generating 3NF in systems of three and more nucleons 
\cite{Deltuva2003a,Deltuva2003b}. In ChEFT \cite{Epelbaum2009}, the 3NF naturally appears in the next-to-next-to-leading order (NNLO). The 3NF proved essential for the proper description of the nuclear matter, among others, for exact calculation of binding energies and low-lying excited states of light nuclei, correct determination of the neutron drip-line and saturation in nuclear matter \cite{Carlson2015, Machleidt2016, Hagen_2016, Piarulli2019, Sammarruca2020}.  

In the proton-deuteron elastic scattering the importance of 3NF contribution to the differential cross section grows fast with energy \cite{Kalantar-Nayestanaki2012}. The residual discrepancy between data and theory, which is significantly reduced when 3NF is taken into account \cite{Witala1998}, also shows a~systematic increase with rising center-of-mass energy. The first attempts at relativistic calculations predict influence on the elastic scattering cross section much too small to be responsible for the observed discrepancy and playing a~role in different kinematic regions 
\cite{Witala2006,Skibinski2006,Witala2011}. The impact of Coulomb interaction  on elastic scattering cross section at medium energies is very small. 

The deuteron breakup in collision with a proton is very interesting as complementary field of studies, since the three-body final state is rich in kinematic configurations revealing diverse sensitivity to specific effects. The sensitivity to 3NF was shown for a~deuteron beam energy of 65 MeV/nucleon and 80 MeV/nucleon \cite{Kistryn2003,Kistryn2005,Parol20}. Moreover, a~predicted huge  influence of the Coulomb interaction \cite{Deltuva2005,Deltuva2006} was experimentally confirmed  \cite{Kistryn2006,Kistryn2013,Ciepal2015}, even at energies as high as 170 MeV/nucleon \cite{WASA2020}. According to the theoretical prediction, the Coulomb effects barely depend on beam energy, and they are very significant for $pp$ pair final state interaction (FSI). The cross-section's description is improved by extending theoretical models with both 3NF and Coulomb interaction \cite{Deltuva2009}. For deuteron beam energy of 130~MeV (65~MeV/nucleon), the improvement of results in the whole phase space of a~detector was observed \cite{Kistryn2013}. 

Both reaction channels in proton-deuteron collisions: elastic scattering and deuteron breakup, are not only a~validation field for the quality of nuclear interaction models, but have been considered for constraining the parameters of ChEFT potentials \cite{Epelbaum2020,Skibinski2023}. So far, there are no calculations including simultaneously all the dynamic ingredients (3NF, Coulomb interactions, relativistic effects), therefore the data base covering large reaction phase space and wide range of beam energies is necessary to single out the particular effects in the range of their maximum visibility. Large detectors combined with modern data acquisition systems facilitate experiments covering large phase space regions. The Cyclotron Center Bronowice (CCB) at Institute of Nuclear Physics PAS, Kraków, provides unique opportunity to continue the studies of few-nucleon systems with the use of proton beams in the energy range of interest for this research, i.e. high enough to expect a significant effect of the 3NF force, and below the threshold for pion production.
At 108 MeV, there exist data for the elastic scattering, in good agreement with the calculations including 3NF \cite{Ermisch2005}. The main goal was to determine the differential cross section for the breakup reaction normalized  relatively to the elastic scattering. 

\section{Experiment}\label{experiment}

The experiment was carried out at CCB with the unpolarized proton beam provided by isochronous cyclotron Proteus C-235. Protons at the kinetic energy of 108 MeV were impinging on the liquid deuterium ($LD_2$) target \footnote{https://experimentsccb.ifj.edu.pl}. The reaction products were measured using the BINA (Big Instrument for Nuclear polarization Analysis) detection system, designed to study the few-nucleon reactions at intermediate energies. 

BINA was previously used for such measurements at KVI, Groningen, The Netherlands, by a~Dutch-Polish collaboration, providing a~rich database for proton-deuteron and deuteron-deuteron collisions \cite{Kistryn2013, Stephan2010,Ciepal_19,Bayat2020,Dadkan2020,Hajar2020,Parol20}. In 2012, the BINA setup was installed at CCB. In addition to mechanical adaptations and a~review of the MWPC electronics etc., the new slow control system was introduced and the decision was made to cut the two middle bars of the $E$-detector in half.

\begin{figure}
    \includegraphics[width=0.3\textwidth]{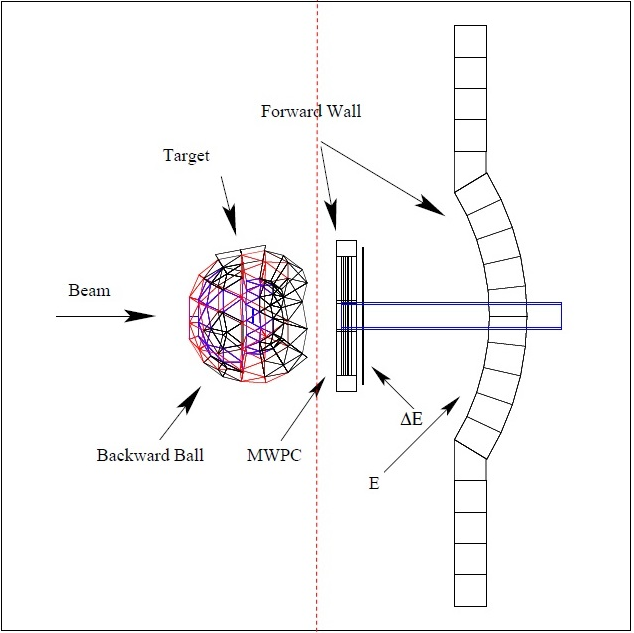}
    \caption{A~schematic side view of the BINA Detection System. The figure is adopted from \cite{Khatri_PhD}}.
    \label{BINA}
\end{figure}

The BINA detector (see Fig.~\ref{BINA}) consist of two parts: a~forward Wall and a~central-backward part - Ball. It allows measurement of charged particle coincidences over a~wide angular range, providing access to an almost full phase-space of few-body scattering reaction. During the first experiment at CCB, dedicated also to the detector commissioning on the new beam line, only the Wall detectors worked optimally, so this paper presents data which were collected in the forward Wall.

\begin{figure}
    \includegraphics[width=0.5\textwidth]{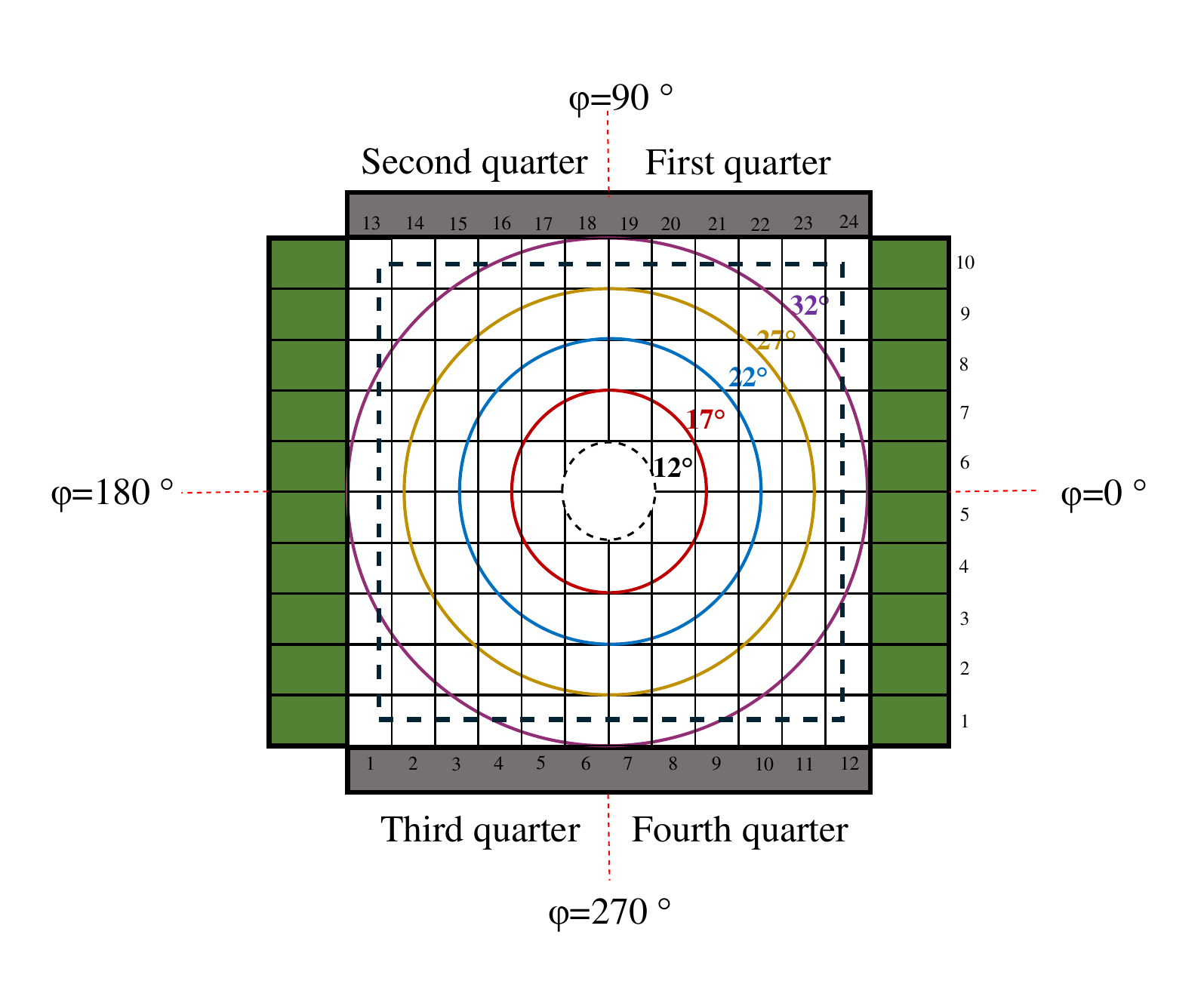}
     \caption{Projection of 24 vertical $\Delta E$ strips and 10 horizontal $E$~bars on the plane perpendicular to the beam axis. The polar angles ($\theta$) are represented by circles. An inner~dashed-line circle represents the beam pipe. Numbering of detector elements and  quarters is given. 
     A rectangle marked with a~dotted line shows approximate limits of the MWPC acceptance.} 
     \label{hodoscope} 
\end{figure}

The Wall part consists of a~Multi-Wire Proportional Chamber (MWPC) and two layers of scintillator hodoscopes forming a~two-dimensional array of  $\Delta E-E$ telescopes for particle identification and setting the trigger conditions (Fig.~\ref{hodoscope}). Particles scattered in the forward direction were measured with the polar angles $\theta$ between $13^\circ - 28^\circ$ in a~whole azimuthal angle $\varphi$. Additionally, the polar angle was extended up to $35^\circ$ at the corners of a~square-shaped MWPC region with only limited coverage of the azimuthal angles. The exact values of the angular acceptance varied slightly between experimental runs due to certain freedom of relative positioning of the Wall-Ball system. The MWPC detector consists of three anode planes with wires oriented vertically (X), horizontally (Y) and diagonally (U). The first layer of the scintillator hodoscopes consists of 24 vertically-placed thin transmission-$\Delta E$ strips. The second one ($E$~detector) is composed of 10 horizontally-placed thick stopping-$E$ bars read from both ends by PMT's. They are forming an arch thus reducing the contribution of so-called "cross-over" events - particles producing a~detectable signal simultaneously in two adjacent bars. Two central scintillators of $E$~detector (No 5 and 6 in Fig.~\ref{hodoscope}) were shaped to make possible passage of the beam pipe, such that only a narrow linking between two sides remained. As mentioned earlier, the decision was taken to cut and optically isolate left and right parts of these detectors. 

The central-backward part of the detector serves also as a~vacuum chamber where the target system is located. During the experiment, we used two kinds of targets. The first one, an aluminum target (Al) with a~thin zinc sulfide ($ZnS$) layer, was prepared to optimize the beam position and was also used during the calibration runs. The second, the liquid deuterium target ($LD_2$), was for studying the reactions of interest. 

The trigger system was designed to detect two kinds of processes: the elastic scattering registered as a~single charged particle and the breakup reaction - coincidences of two protons. The individual prescaling factors were applied to obtain representative samples of both event types. For more information on the electronics, readout and data acquisition systems, see Ref.~\cite{Stephan_hab}. The triggers were determined by logic conditions based on: signals from $\Delta E$ detectors, left PMT's of $E$~detector ($E_L$), and right PMT's of $E$~detector ($E_R$). The logic conditions of the triggers are presented in Table~\ref{triggers_fig}. Trigger T1 was a~minimum bias trigger, with at least one charged particle in the whole detection setup (single particle). The T2 trigger was based on the multiplicity hits, and it was used to select the coincidences of two particles registered in the Wall detector (Wall-Wall coincidence).

\begin{table}
\caption{The logic conditions of the triggers. MULT denotes condition on hit multiplicity.} 
\centering
\begin{tabular}{p{0.15\linewidth} | p{0.65\linewidth} | p{0.15\linewidth}}
    \hline
    \textbf{ } & \textbf{} & \textbf{Down-} \\
    \textbf{Symbol} & \textbf{Logic notation} & \textbf{scaling factor} \\
    \hline
    \hline
    \textbf{T1} &  (OR $E_L$) OR (OR $E_R$) OR (OR $\Delta E$) &  $2^4$ \\
    \hline
    \textbf{T2} & (MULT $E_L \ge 2$) OR (MULT $E_R \ge 2$) & $2^0$ \\
    \hline
\end{tabular}   
        
\label{triggers_fig}
\end{table}

\section{Data analysis} \label{analysis}

Data analysis started with selection of events time-correlated with the trigger and identification of reaction channels of interest. 
The next steps included energy calibration, determination of detection efficiency and determination of luminosity for the cross section normalization.

\subsection{Event selection and reconstruction} \label{track}
    
\begin{figure}
        \centering
        \includegraphics[width=0.3\textwidth]{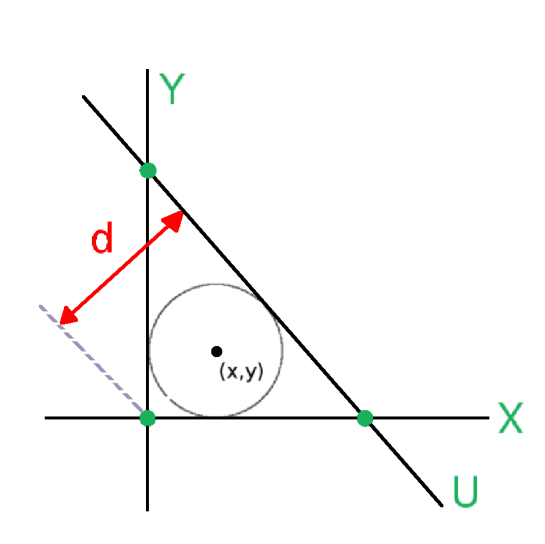}
        \caption{A graphic represents three hit wires X,~Y,~and U,~which are the basis of the track reconstruction procedure. Center of the circle inscribed in the XYU triangle represents position assigned to the particle track.}
        \label{track_rec}
\end{figure}

The track reconstruction was based on hits in the MWPC planes. The first step of the track reconstruction procedure (described in details in Ref.~\cite{Parol20}) was to project the coordinates of hits in X~and U~planes onto a~central Y~plane. The point of passage of the particle was determined as the center of the circle inscribed in the right angled triangle formed by the hit wires (see Fig.~\ref{track_rec}). The hits were treated as correlated (produced by the same particle) when the distance $d$~between the XY intersection and the active wire U~was less than 7~mm. Assuming the reaction point in the target center, we calculated polar ($\theta$)~and azimuthal ($\varphi$) scattering angles in the laboratory frame. 

\begin{figure}
        \centering
        \includegraphics[width=0.45\textwidth]{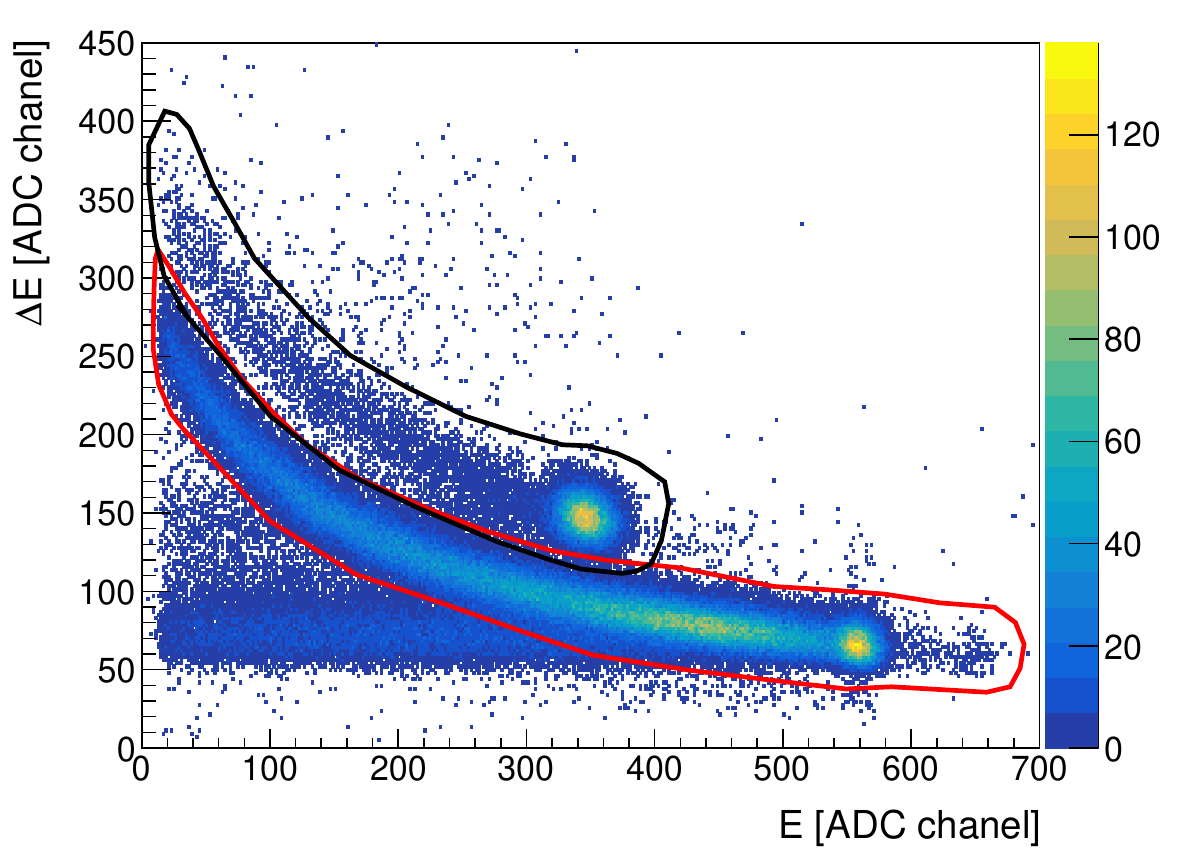}
        \caption{Two-dimensional spectrum built for one $\Delta E-E$ telescope ($\Delta E=2$, $E=0$); good separation of protons (lower band; red gate) and deuterons (upper spot, black gate) is visible.}
        \label{PID_fig}
\end{figure}

The analysis focused only
on tracks with MWPC hits matched with the $\Delta E$ and $E$~hodoscopes. The particle identification (PID) of both particle types was based on the $\Delta E-E$ technique and an example of such a~identification spectrum is presented in Fig.~\ref{PID_fig}. Proton and deuteron distributions can be well distinguished, thus simple graphical cuts (red and black contours in Fig.~\ref{PID_fig}) were implemented. The gates were wide enough to avoid a~significant loss of particles, and a~slight overlap of them was allowed, see Fig.~\ref{PID_fig}. 

In order to perform the energy calibration, dedicated runs were carried out with proton beams at energies of 70, 83, 97, 108 and 120 MeV impinging on an aluminum target. The registered particles were sorted by the side \textit{s}~(\textit{s = right / left}), the detector number \textit{n}~(\textit{n = 1, 2, ..., 10}), and polar angle $\theta$~binned by 4~degrees. The signals $c_1$ and $c_2$ registered by photomultiplier tubes on both ends of each $E$~detector were combined in the geometrical average ($C=\sqrt{c_{1}\cdot c_{2}}$) in order to suppress the effects of light attenuation along the bar. Due to the cutting of the two central scintillators, in their case the geometrical average was replaced by a~signal of one side ($C=c_{1}$ or $C=c_{2}$). The spectra of protons scattered on the Al target, were analyzed for a~chosen set of parameters: beam energies, $s,~n$ and $\theta$.~On the basis of the Monte Carlo simulation performed in Geant4, the actual energy deposited by protons, $E_{dep}$, was determined. The relation between $C$~and $E_{dep}$ is described by  linear calibration above approximately 50 MeV. Below this value, the light quenching effect is expected due to the large stopping power~\cite{Quench1, Quench2}. The calibration curve fitted to the entire range of  $C$~is described by the non-linear equation \cite{Lobejko_PhD}:
\begin{equation}\label{en_cal}
        E_{dep} \ (s,n,\theta) = a \ (s,n,\theta) \cdot C + b \ (s,n,\theta) \cdot \sqrt{C};
    \end{equation}
where: $a \ (s,n,\theta)$ and $b \ (s,n,\theta)$ denote the fit parameters. 
  
Having reconstructed particle trajectory, a~series of simulations were performed to determine the particle's energy at the point of reaction. Finally, we obtained the initial ($E_{initial}$) versus deposited energy ($E_{dep}$) spectra for each angular bin, for details see Ref.~\cite{Lobejko_PhD}.

\subsection{Corrections for detection efficiency and hadronic interactions} \label{eff}

\begin{figure}[]
        \centering
        \includegraphics[width=.4\textwidth]{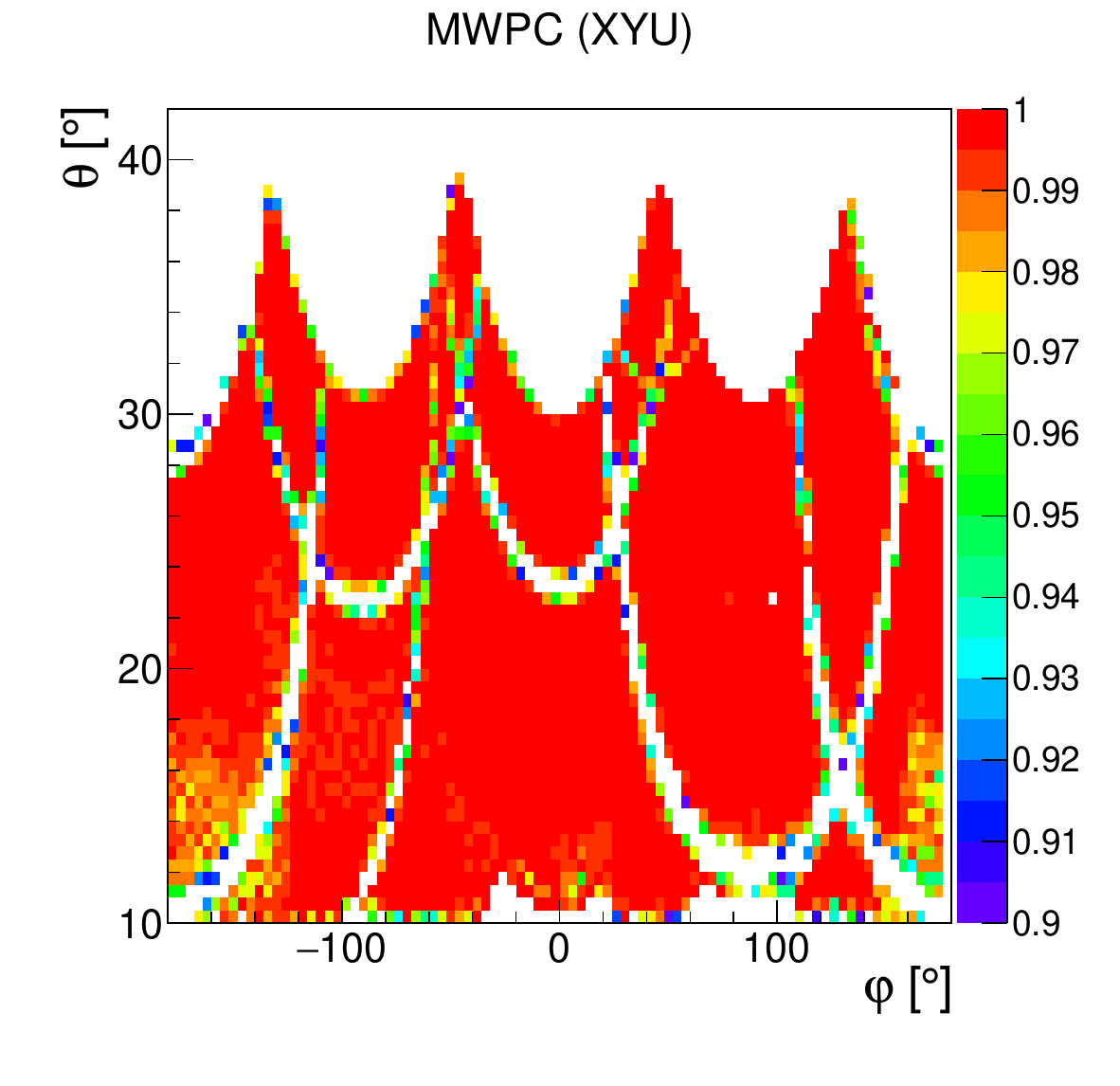} \\
        \includegraphics[width=.4\textwidth]{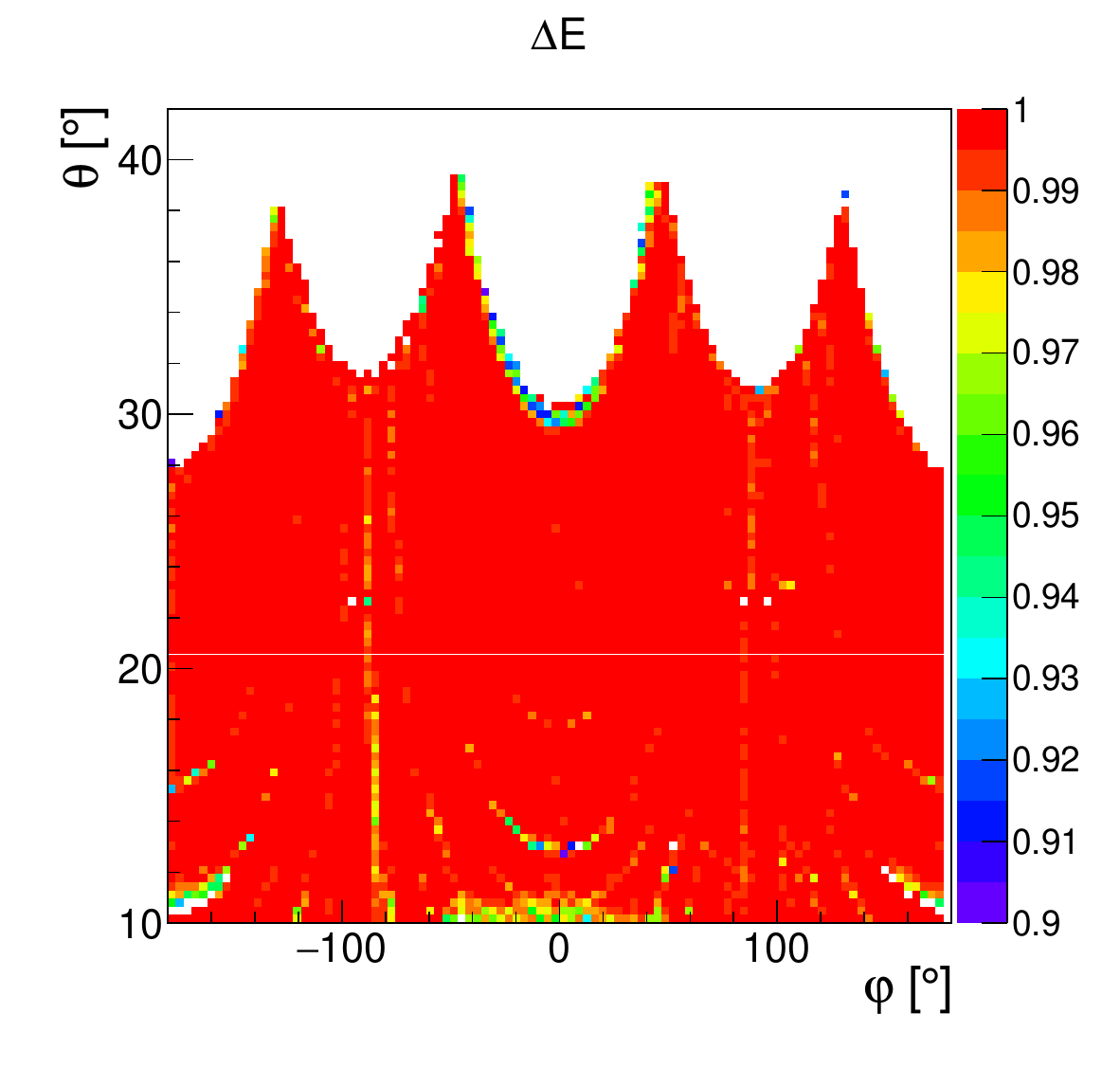}
        \caption{The $2$-dimensional efficiency maps  for MWPC (\textit{upper panel}) and $\Delta E$ detector (\textit{lower panel}).}
        \label{fig_eff_MWPC}
\end{figure}

In the case of elastic scattering, a~single particle was recorded in the Wall. Therefore, $1$-dimensional efficiency as a~function of the $\theta$~angle was used. The breakup reaction is reconstructed from coincidences of two protons and thus characterized by four non-separable variables: two polar and two azimuthal angles. Consequently, a~$2$-dimensional ($\theta$,~$\varphi$) efficiency map was constructed for this purpose and employed for the event-by-event correction. In addition, we considered the configuration efficiency of the system related to the geometrical acceptance for registering two coincident protons in separate detector elements. The total efficiency is defined as a~product:
\begin{equation}\label{eff_wall} 
   \varepsilon_{total}    = \varepsilon_{MWPC} \cdot \varepsilon_{\Delta E} \cdot \varepsilon_{E}  \cdot {\varepsilon}_{conf},
    \end{equation}
where: 
$\varepsilon_{MWPC}, \ \varepsilon_{\Delta E}, \  \varepsilon_{E}, \  \varepsilon_{conf}$ are the efficiencies of the Multiwire Proportional Chamber, $\Delta E$, $E$~detectors, and configurational efficiency (only in the case of the breakup reaction), respectively. 
To avoid ambiguity, only single-particle events recorded in the $\Delta E$ and $E$ hodoscopes were considered to construct the map of MWPC efficiency.  
The MWPC efficiency, see Fig.~\ref{fig_eff_MWPC}, \textit{upper panel}, is a~product of efficiencies of all its three planes: $\varepsilon_{MWPC} \ (\theta,\varphi) = \varepsilon_{x} \ (\theta,\varphi)  \cdot \varepsilon_{y} \ (\theta,\varphi)  \cdot \varepsilon_{u} \ (\theta,\varphi)$, calculated as follows:
\begin{equation}\label{eff_X}
        \varepsilon_{x} \ (\theta,\varphi) = \frac {N_{xyu} \ (\theta,\varphi)}  {N_{yu} \ (\theta,\varphi)} 
    \end{equation}    
where: 
\begin{itemize}
    \item $N_{xyu} \ (\theta,\varphi)$ is a~number of counts registered in the selected angular segment, satisfying the condition of one hit (or one cluster of hits) per the $X$, $Y$, and $U$~planes;
    \item $N_{yu} \ (\theta,\varphi)$ is a~number of counts registered in the same angular segment with at least one hit (or cluster) per the $Y$ and $U$~planes.
\end{itemize}

The $\Delta E$ efficiency (see Fig.~\ref{fig_eff_MWPC}, \textit{lower panel}) was calculated similarly, by counting the ratio of the number ($N_{\Delta E}$) of particles reconstructed based on signals from MWPC, $\Delta E$ and $E$, to those ($N_{all}$) registered regardless of the presence of information from $\Delta E$ : 
\begin{equation}\label{eff_dE}
        \varepsilon_{\Delta E} \ (\theta,\varphi) = \frac {N_{\Delta E} \ (\theta,\varphi)}  {N_{all} \ (\theta,\varphi)} 
    \end{equation}

Scintillator elements of the $E$~detector are thick enough to stop all the charged particles emitted at the beam energy of 108 MeV. The gaps between the scintillator blocks have the thickness of the foil wrapped around the detectors and their influence on efficiency was considered to be negligible.

The proper registration of two charged particles in the BINA Wall required two reconstructed hits in all the MWPC planes and signals from two various elements of $\Delta E$ detector, and two elements of the $E$~detector. The probability of losses due to two protons impinging the same detector element strongly depends on the angular configuration. There were also event losses due to inefficiency of the coincidence trigger. The coincidence trigger (T2) condition described in Table.~\ref{triggers_fig} requires two signals in $E$~detector's left or right side. Since the central detectors were cut in the middle, the PMT's signal from those detectors was always registered from one (left or right) side only and for certain event geometries the trigger condition was not fulfilled. All these effects were accounted by the total configurational efficiency obtained by analysis of Geant4 Monte Carlo simulations. The efficiency was calculated for each analyzed breakup configuration as a~ratio of the number of events with complete information from all the detectors and the geometry compatible with the T2 trigger condition ($N_{g}$) to the total number of simulated events ($N_{all}$):
\begin{equation} \label{e_conf}
    \boldsymbol{\varepsilon}_{conf}(\theta_1, \theta_2, \varphi_{12})=\frac{N_g (\Delta E, E)}{N_{all} (\Delta E, E)}.
\end{equation}

\begin{figure}[]
        \centering
        \includegraphics[width=0.45\textwidth]{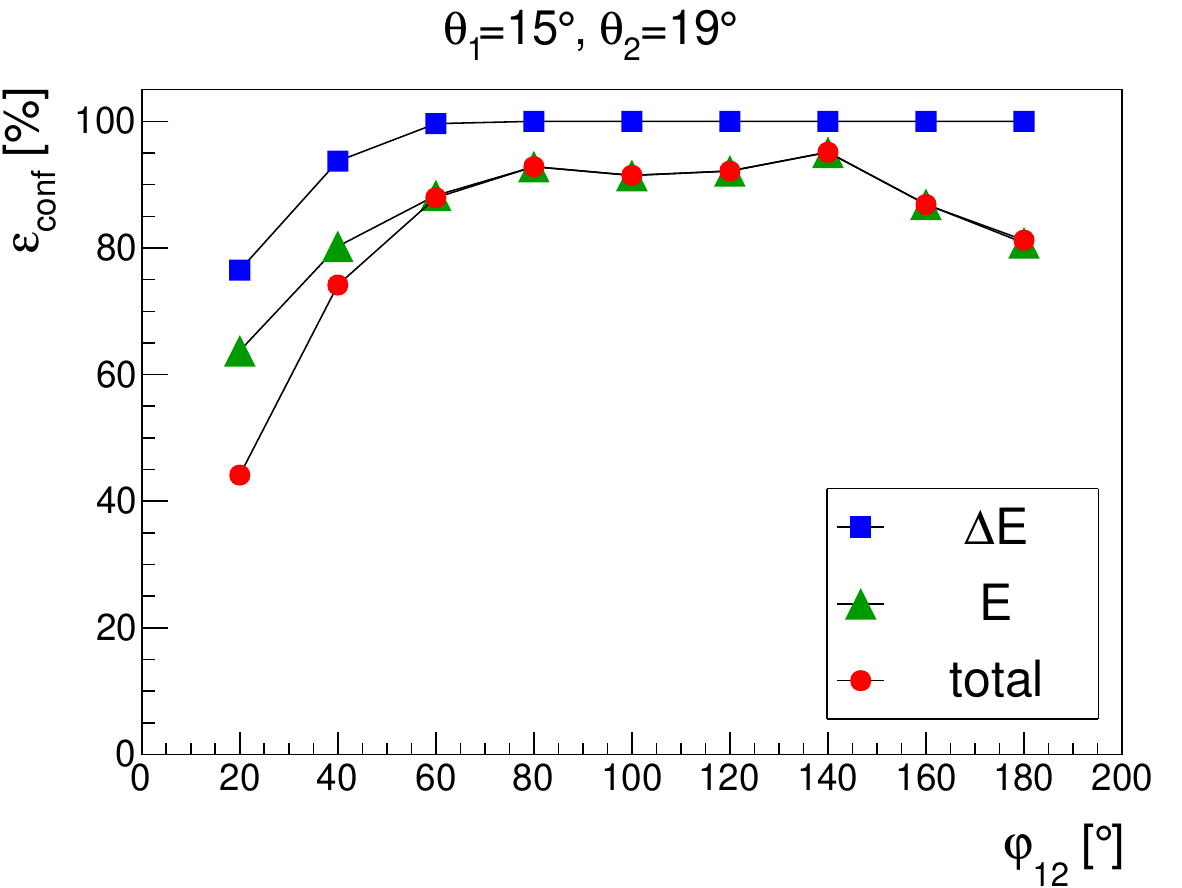}
        \caption{
        Configurational efficiency for selected $\theta_1$, $\theta_2$ and all $\varphi_{12}$,  obtained in simulation. The blue squares and green triangles refer to the efficiency calculated separately for $E$~and $\Delta E$. The red points represent the total configurational efficiency for the BINA detection setup.}
        \label{wyd_konf}
    \end{figure}

Three sets of results obtained for one chosen combination of polar angles $\theta_{1}$ and $\theta_{2}$ are presented in function of $\varphi_{12}$ in Fig.~\ref{wyd_konf}. As expected, the configurations with the smallest relative angles $\varphi_{12}$ are particularly prone to configurational efficiency losses. The coplanar ones ($\varphi_{12} = 180^{\circ}$) are affected by the trigger inefficiency.

A~set of simulations was carried out to calculate the fraction of particles undergoing hadronic interaction~\cite{Ciepal_19}. For deuterons, losses due to hadronic interactions have also been estimated from the experimental data~\cite{Lobejko_2023, Lobejko_PhD}. The difference between the measured and simulated contribution of hadronic interactions is of around 12\% relatively, corresponding to a 2\% systematic error of the integrated luminosity.

\subsection{Determination of integrated luminosity} \label{lum_det}

The integrated luminosity was calculated based on the number of the elastically scattered particles $N^{el}(\theta)$, and the known elastic scattering cross section $\sigma \ ^{el}_{LAB}\ (\theta)$ \cite{Ermisch2005}. Elastically scattered particles, protons or deuterons, were registered in the forward Wall as single tracks since their coincident particles were emitted towards Ball. In the proton energy spectra, elastically scattered protons were contaminated with protons from the breakup reaction, and there was also small contribution of protons scattered from the target frame. Therefore, the luminosity was determined based on the analysis of much cleaner deuteron spectra.

\begin{figure}[]
        \centering
        \includegraphics[width=.45\textwidth]{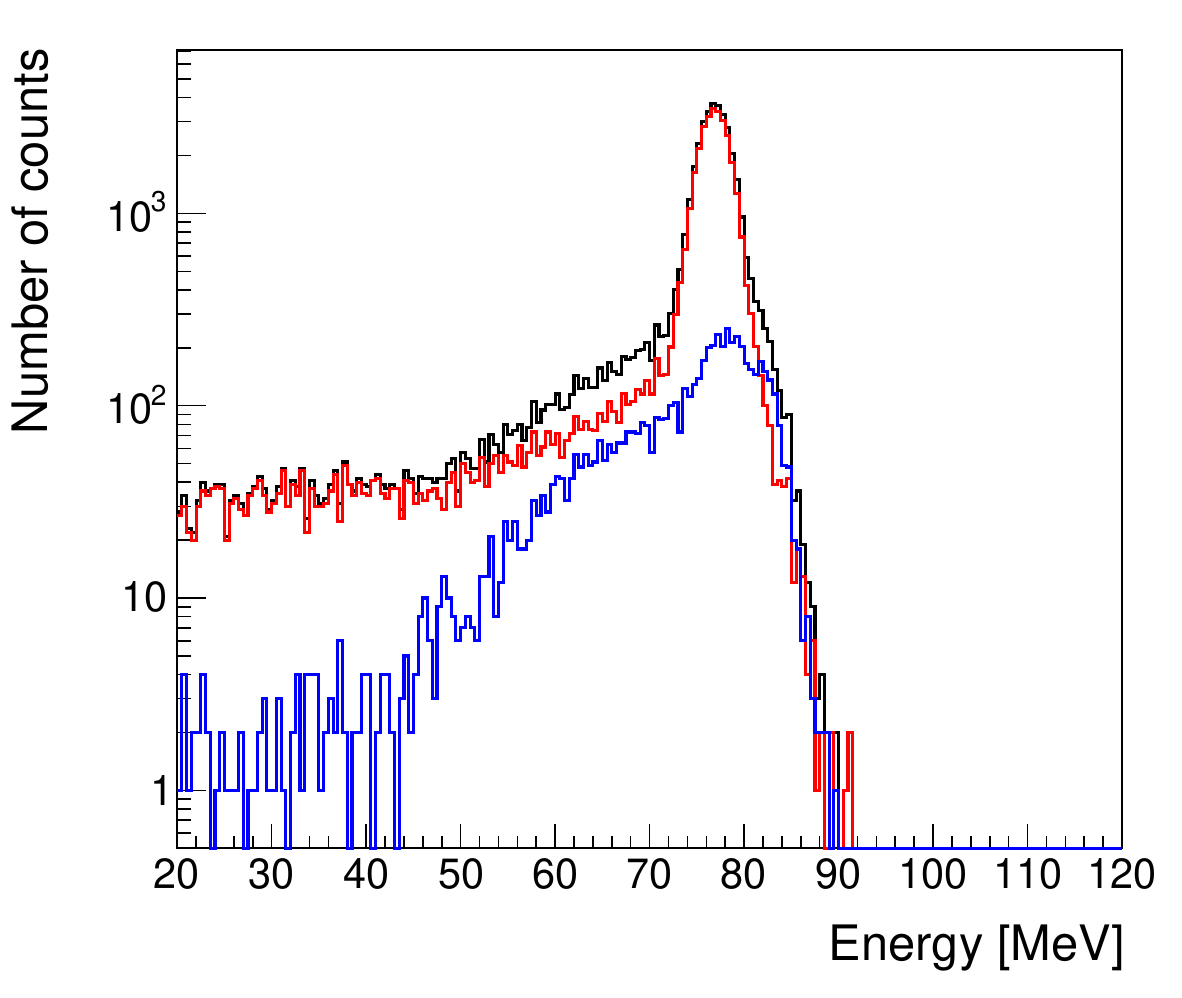}
        \caption{
        Spectra of deuterons registered at $\theta=20^\circ$ obtained with two PID cuts: a~broad gate (black line) and a~narrow gate (red line); blue line represents a difference of these spectra.}
        \label{hadr_fig}
    \end{figure}

Deuterons were sorted according to their polar angles. The energy spectrum of deuterons  elastically scattered at $\theta=20^\circ\pm 0.5^{\circ}$ is presented in Fig.~\ref{hadr_fig}. Narrowing the deuteron PID gate (so as to eliminate its overlap with the proton gate, see Fig.~\ref{PID_fig}) we observed a~significant reduction of the background, thus the protons were the main contributor. This is confirmed by the presence of a~peak in the difference between the spectra for the wider and narrower gates (blue line in Fig.~\ref{hadr_fig}), which extends above the deuteron energies. After subtracting the background  approximated by straight lines, the sums under the peaks for both gate widths are very similar, with slightly larger values for the wide gate. Finally, for the data analysis, a~wide gate was chosen to avoid deuteron losses.  
The  resulting $N^{el}(\theta)$ values were corrected for the detection efficiency $\varepsilon_{total}(\theta)$, hadronic interactions and a~solid angle $\Delta\Omega$. Due to the square shape of our detection setup, starting from a~certain theta angle the limited azimuthal acceptance should be taken into account. For the two largest analysed polar angles, limits on the MWPC corners, $\varphi=45^\circ \pm 10^\circ, 135^\circ \pm 10^\circ, 225^\circ \pm 10^\circ, 315^\circ \pm 10^\circ$, were applied.

\begin{figure}[]
        \centering
        \includegraphics[width=0.45\textwidth]{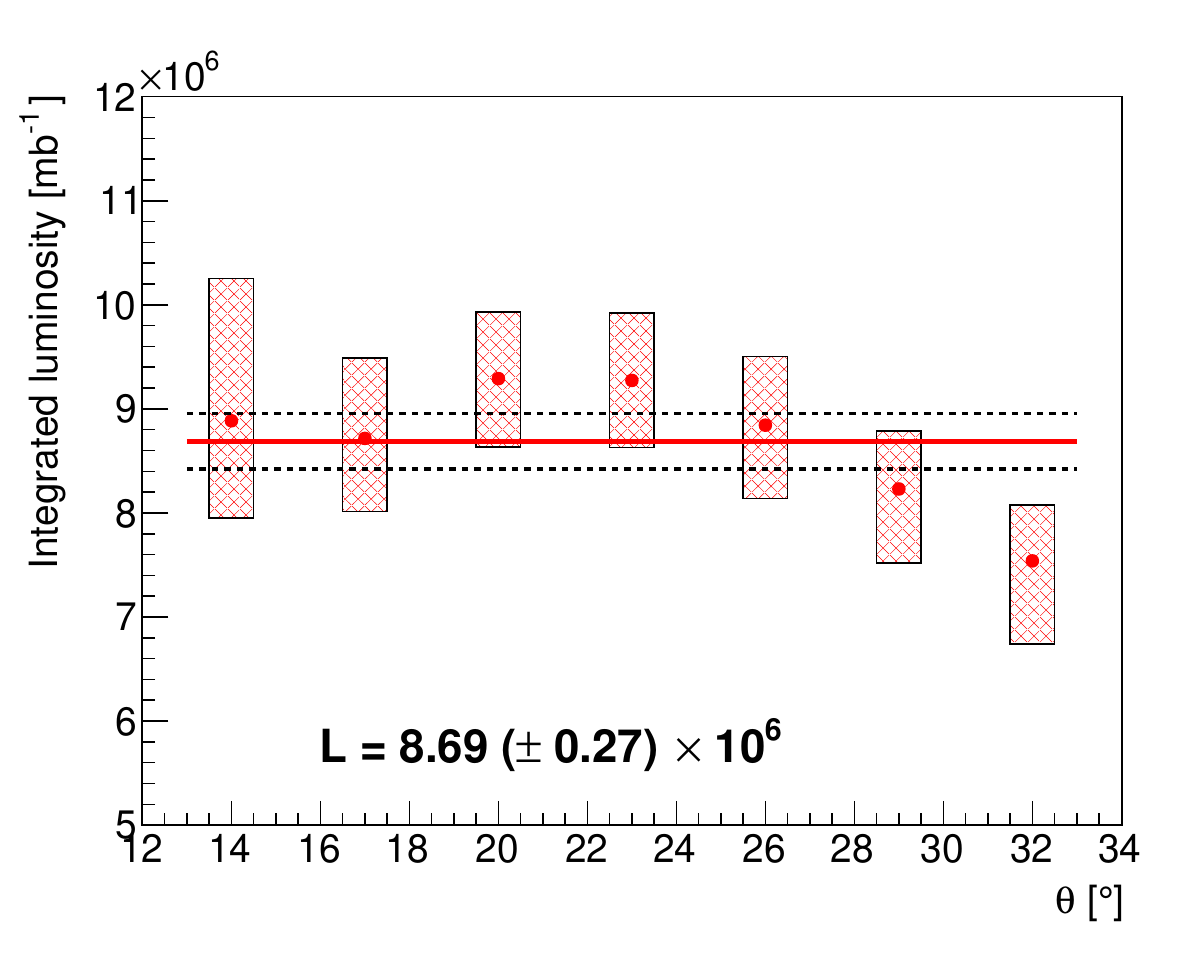}
        \caption{The luminosity value obtained for the set of $\theta$~angles, calculated as a~mean weighted by a~systematic error represented by the hatched boxes (more details in text). The statistical errors are smaller than the point's size.}
        \label{lum_pic}
\end{figure}

We normalized our data to the differential cross section data for \textit{pd} elastic scattering at 108 MeV that were measured at KVI with the use of the Big Bite Spectrometer~\cite{Ermisch2005}. The cross-section error given in Ref.~\cite{Ermisch2005} ranges from $4.4\%$ for the highest $\theta$~angles to $6.5\%$ for the lower ones. The second source of systematic uncertainty is related to our measurement and analysis. As a~measure of it, we took the variations of the count's rates in the four quarters of the detector \cite{Lobejko_PhD}. The total systematic uncertainty varies between $6.8\%$ and 15.4\% depending on the $\theta$ angle. The statistical errors are minor and range between $0.78-0.85\%$. Fig.~\ref{lum_pic} shows the resulting luminosity values with their systematic uncertainties. 
The average luminosity is shown in Fig.~\ref{lum_pic} as a~red horizontal line. The points in the $\theta$ range between $14^{\circ}-29^{\circ}$ are consistent within the systematic uncertainty represented by the black dashed lines, while the distance of the last point corresponding to $\theta=32^{\circ}$ goes beyond the range of error. It is on the limit of the geometrical acceptance of the Wall detector, but the acceptance losses have been taken into account, thus indicating small shape difference between our results and the KVI data~\cite{Ermisch2005}. The influence of this point on the average value is insignificant (within systematic uncertainty of the average).

\section{Determination of experimental differential cross section} \label{diff_br}

The differential cross section for the $^2$H(p,pp)n reaction was analyzed for a~set of kinematic configurations within the angular acceptance of the Wall. The breakup reaction kinematics is determined by the momenta of protons $\vec{p_{1}}$ and $\vec{p_{2}}$. The geometrical configuration is defined by polar angles $\theta_{1}, \theta_{2}$ of the momentum vectors and their relative azimuthal angle $\varphi_{12}$. The events identified as proton-proton coincidences were corrected by efficiency factors and sorted according to their angular configuration ($\theta_{1}, \theta_{2}, \varphi_{12}$). The order of protons in the case of symmetric configurations was randomized, otherwise, the proton with a~lower polar angle ($\theta$) was chosen as the first one. The range of the polar angles $\theta_1$ and $\theta_2$ from $13^{\circ}$ to $33^{\circ}$ has been divided into 4-degree wide bins, and the relative azimuthal angle $\varphi_{12}$ was sorted into bins of 20 degrees. Configurations characterized with the lowest azimuthal angles were a~subject to large losses of configurational efficiency, thus they required a~specific approach to the analysis. The results presented here are limited to $\varphi_{12}\geq50^{\circ}$. We also considered the acceptance losses due to the square shape of the detector. Similarly to the elastic scattering, safe limits on the azimuthal angle: $\pm 10^\circ$ around each diagonal were set.
Since the bins cover the range $\theta \in(\theta_c - 2^{\circ},\theta_c + 2^{\circ})$ around the central $\theta_c$ value, the acceptance limits  already affected the configuration with $\theta_c=27^{\circ}$. Since setting the limits on azimuthal angles of both protons would also affect $\varphi_{12}$, the configurations where both the coincident protons fulfilled the condition $\theta_c\geq27^{\circ}$ were excluded from the analysis.  
    \begin{figure}[]
        \centering
        \includegraphics[width=0.4\textwidth]{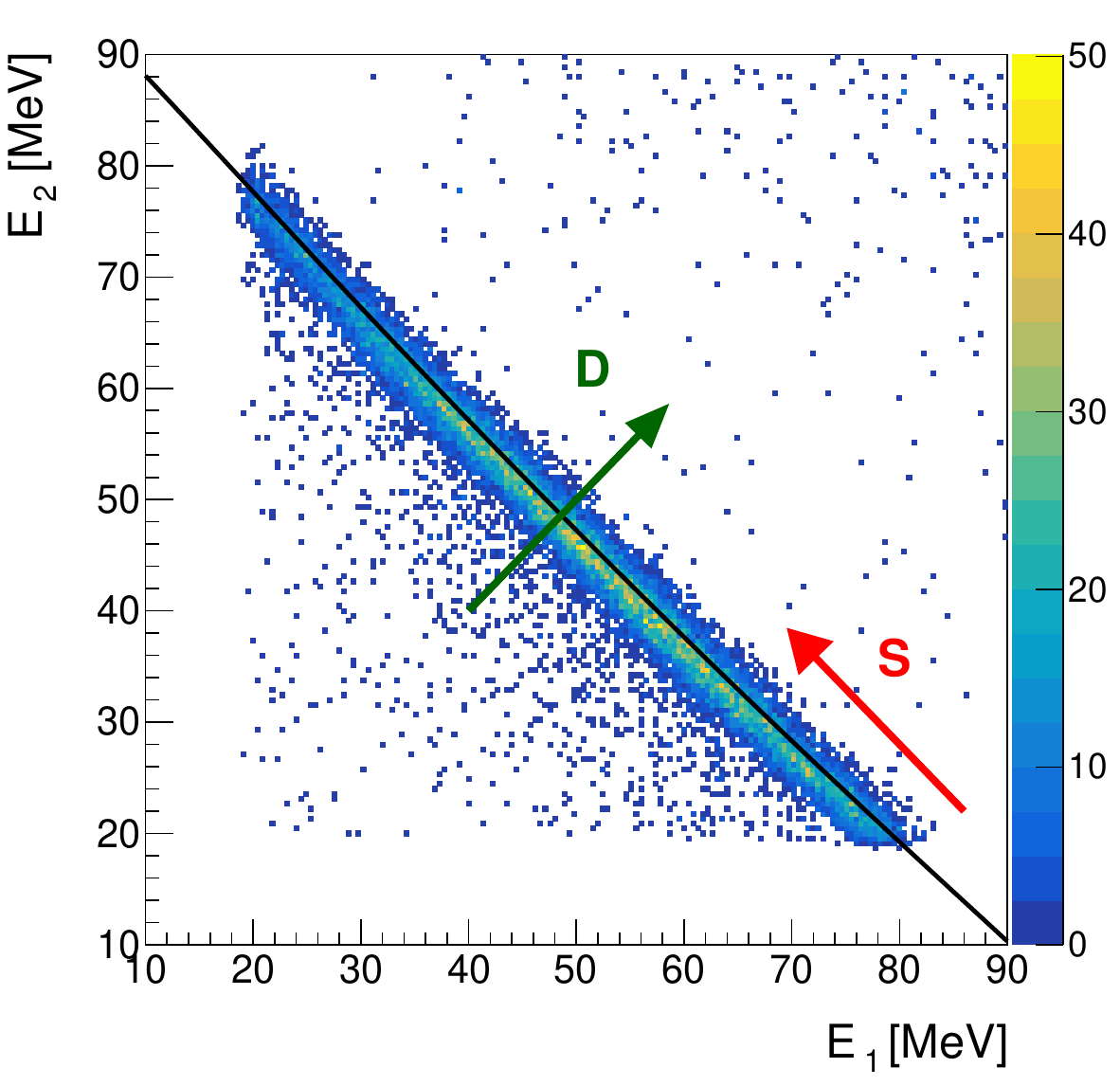}
        \includegraphics[width=0.4\textwidth]{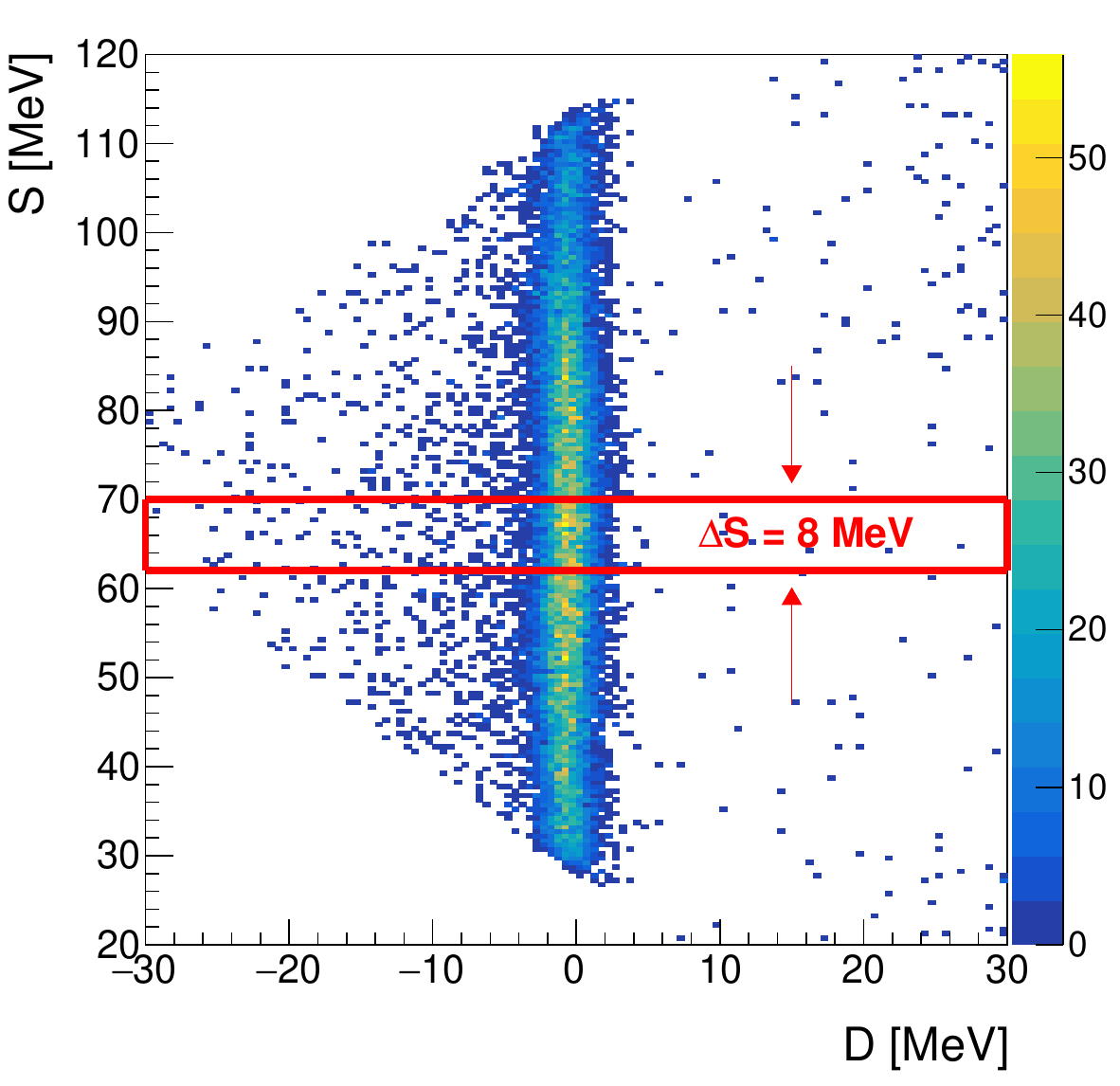}
        \caption{\textit{Upper Panel}: $E_{2}$ vs.~$E_{1}$~spectrum of the proton-proton coincidences obtained for a~selected angular configuration $\theta_{1} = 15^\circ, \theta_{2} = 19^\circ, \varphi_{12} = 160^\circ$. The calculated kinematics is presented as a~solid, black line. The arrows show new variables: distance ($D$)~and arc length ($S$), described in the text. \textit{Lower Panel}: The $E_{2}$ vs. $E_{1}$~spectrum transformed into $S$ vs. $D$~variables. The red box represents one slice of $S$~equal to $\Delta S=$ 8~MeV, projected onto the $D$~axis in the next step.}
        \label{S_D}
    \end{figure}
    
Data correctly classified to the angular configuration should group around the  kinematics in the $E_2$ vs.~$E_1$ plane of proton energies (see Fig.~\ref{S_D}, \textit{upper panel}). Such a~spectrum can be transformed into two other kinematic variables. The $S$~variable denotes the arc length value measured along the kinematic curve (see Fig.~\ref{S_D}, \textit{upper panel}) with the starting point $S=0$~chosen arbitrarily at the point where the energy of the second proton reaches the minimum. In this analysis, the kinematic curve was sliced along its length into the segments of $\Delta S=$ 8~MeV. The $D$~variable is the distance of each data point from the theoretical kinematic curve in the $E_{2}-E_{1}$ plane. The example of the $E_{2}$ vs. $E_{1}$ histogram transformed into the $S$~vs. $D$~spectrum is presented in Fig.~\ref{S_D}, \textit{lower panel}. For each $\Delta S$ slice, the events were projected into  the $D$~axis to perform the background subtraction. The background was estimated by a~linear function between limits of integration ($E_{a}, E_{b}$) corresponding to a~distance of $\pm3\sigma$ from the peak position. Events below the linear function were subtracted. The number of events, $N_{br}(\theta_{1},\theta_{2},\varphi_{12},S)$ was summed up within the $S$-bin and normalized to the integrated luminosity. For each angular configuration, the $S$~distribution of the cross section was obtained, see examples in Fig.~\ref{br_cross_sec_rys}.

    \section{Experimental uncertainties and data consistency checks}

The  experimental uncertainties are summarized in Table \ref{errors_table}. The statistical uncertainties were calculated for each point separately. The systematic errors are estimated as global uncertainties, i.e. the same for all experimental points except for the uncertainty on configurational efficiency, which is calculated for each angular configuration separately.

\begin{table}[]
\caption{Summary of 
uncertainties.}  
\centering
\begin{tabular}{|l|c|}
    \hline
    \hline
     \multirow{2}*{\textbf{Sources of errors}} & \textbf{The impact on breakup} \\ & \textbf{cross section [\%]} \\
      \hline
      \hline
    \textbf{Statistical uncertainties} & $2.3-11.1\%$  \\
     \hline
    \textbf{Total systematic error} & $3.9-8.3\%$  \\
    \hline
    1. Normalization & $3.5\%$  \\

    2. Particle identification & $1\%$  \\
    
   3. Configurational efficiency & $0.01-6.8\%$  \\
    
   4. Energy calibration &  \\ 
   \hspace{5mm}+ angle reconstruction &   \\ 
   \hspace{5mm}+ detector efficiency & $1\%$  \\
    
   5. Trigger efficiency & $-0\%, +3\%$  \\

   6. Hadronic interactions & $1\%$  \\
    \hline  
   
\end{tabular}   
       \label{errors_table}
\end{table}

The potential sources of the systematic uncertainties and estimation of their effect on the cross section value are briefly described below.

\begin{enumerate}

    \item Normalization

The systematic error of luminosity value is 3.5\%, see discussion in Sec.~\ref{lum_det}.

    \item Particle identification
    
To estimate the impact of PID cuts we analyzed the events regardless of the particle type. The distribution of breakup events for both cases (with the presence (\textit{left panel}) and absence (\textit{right panel}) of an identification gate) for the same $S$-variable and kinematic configuration are presented in Fig.~\ref{pid_syst}. The slight increase of the cross section value calculated without the identification gate is mainly caused by a~higher number of random coincidences and the contribution of hadronic events. After background subtraction, the impact of the identification procedure on the cross section value was estimated at 1\%.

\begin{figure}[]
        \centering
        \includegraphics[width=0.23\textwidth]{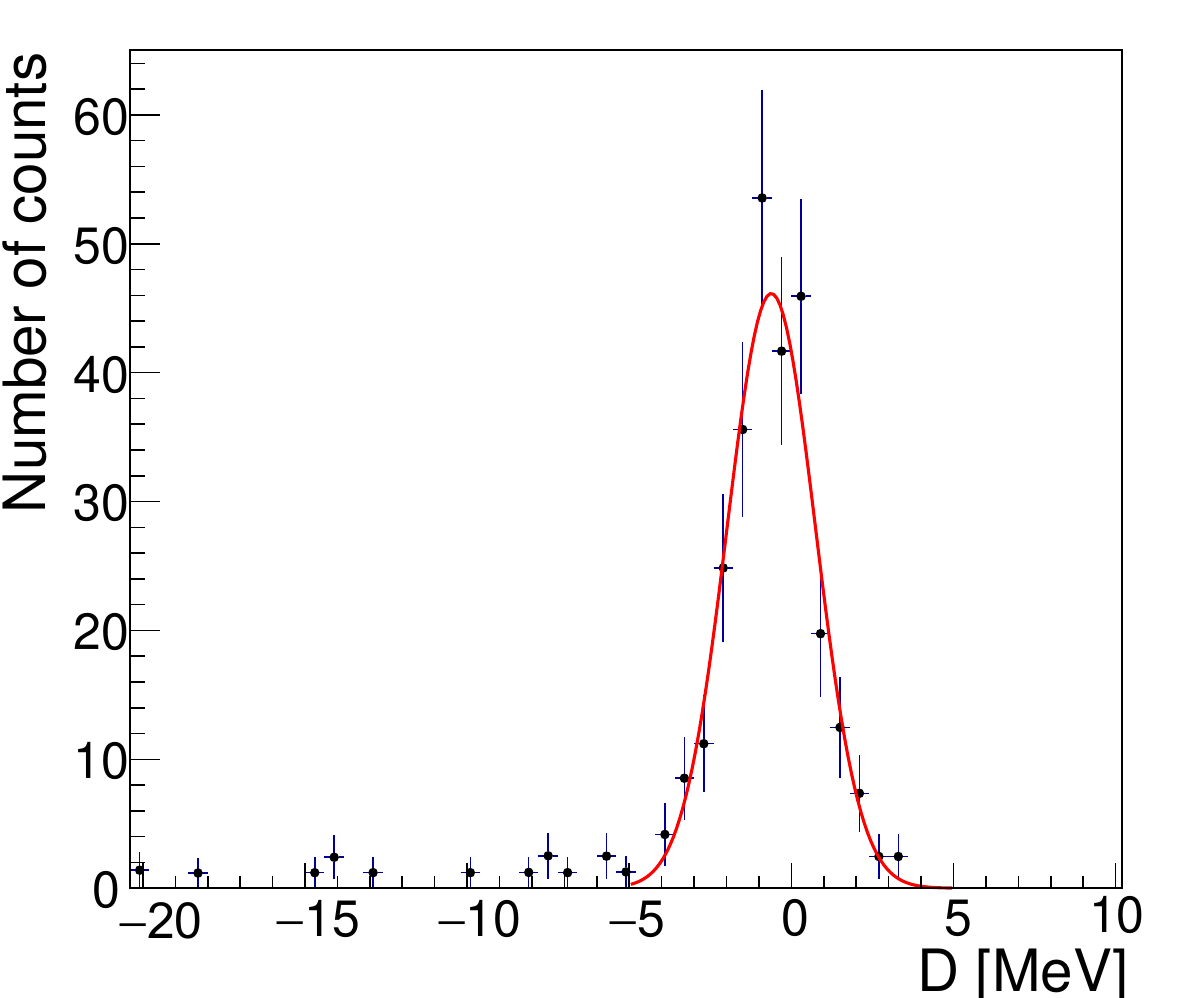}
        \includegraphics[width=0.23\textwidth]{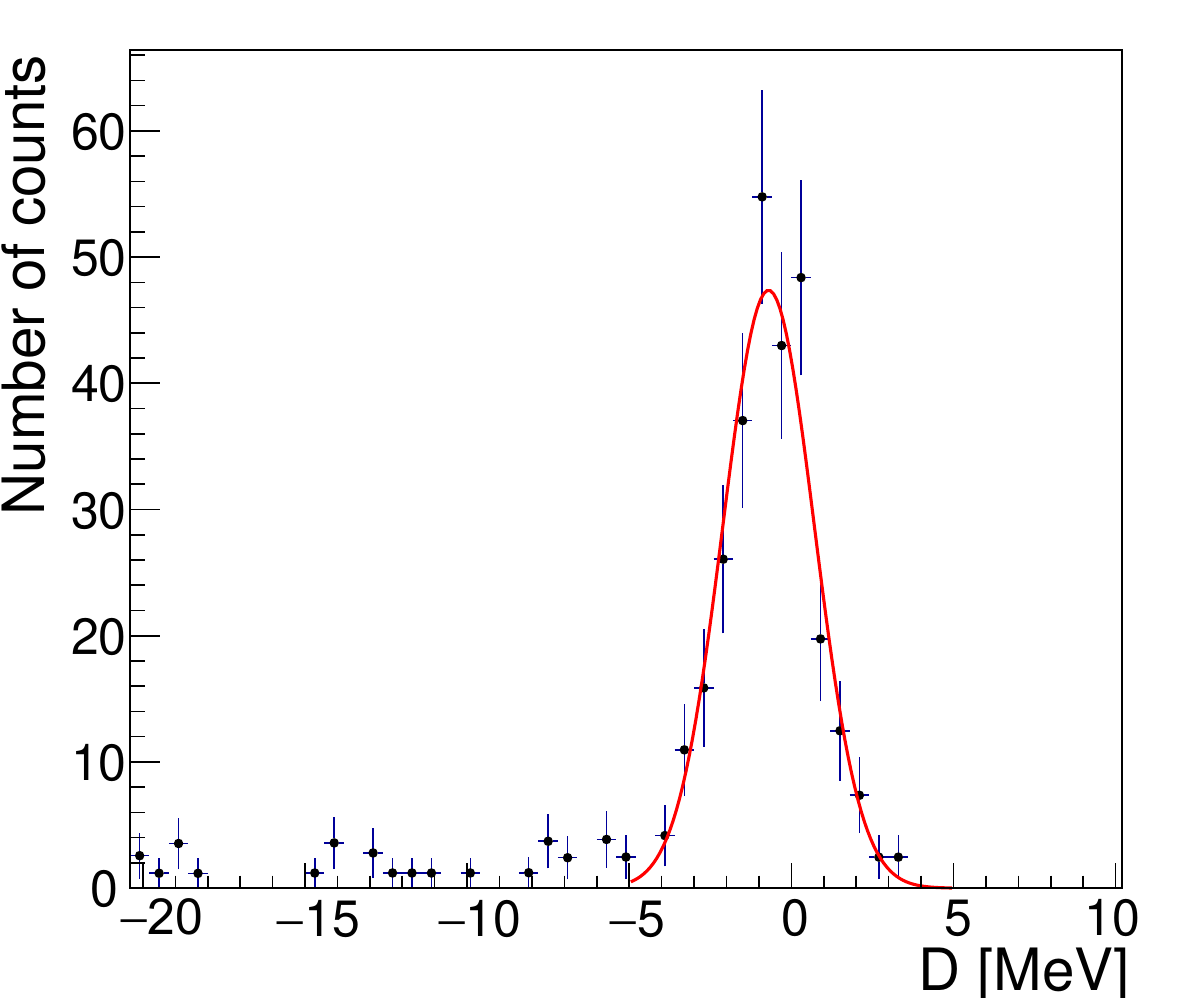}
        \caption{Example of $D$~distribution of proton pairs from the breakup reaction obtained with (\textit{left panel}) and without (\textit{right panel}) the PID cut; data correspond to one $\Delta S$ bin of the angular configuration $\theta_1=18^{\circ}, \theta_2=20^{\circ}, \varphi_{12}=120^{\circ}$.}
        \label{pid_syst}
    \end{figure}

    \item Configurational efficiency
     
Configurational efficiency is closely related to the geometry of the detection system. The geometry of the Wall was studied on the basis of events particularly sensitive to its details, registered on the edges of $E$-detector bars, see Ref.~\cite{Lobejko_2024_FewBody}. According to the results, the simulated setup was modified by introducing a~1 degree inclination of the $E$~detector. The simulation results obtained in both approaches (for nominal and modified geometries) for one combination of $\theta_1=15^{\circ}$ and $\theta_2=19^{\circ}$ angles are compared in Fig.~\ref{conf_eff_syst}. For all the breakup configurations the relative difference varies between 0.01\% and 6.8\%, strongly depending on the relative azimuthal angle $\varphi_{12}$. Conservatively, we treat this difference as the systematic uncertainty of the configurational  efficiency and apply individually to each configuration.

\begin{figure}[]
        \centering
        \includegraphics[width=0.45\textwidth]{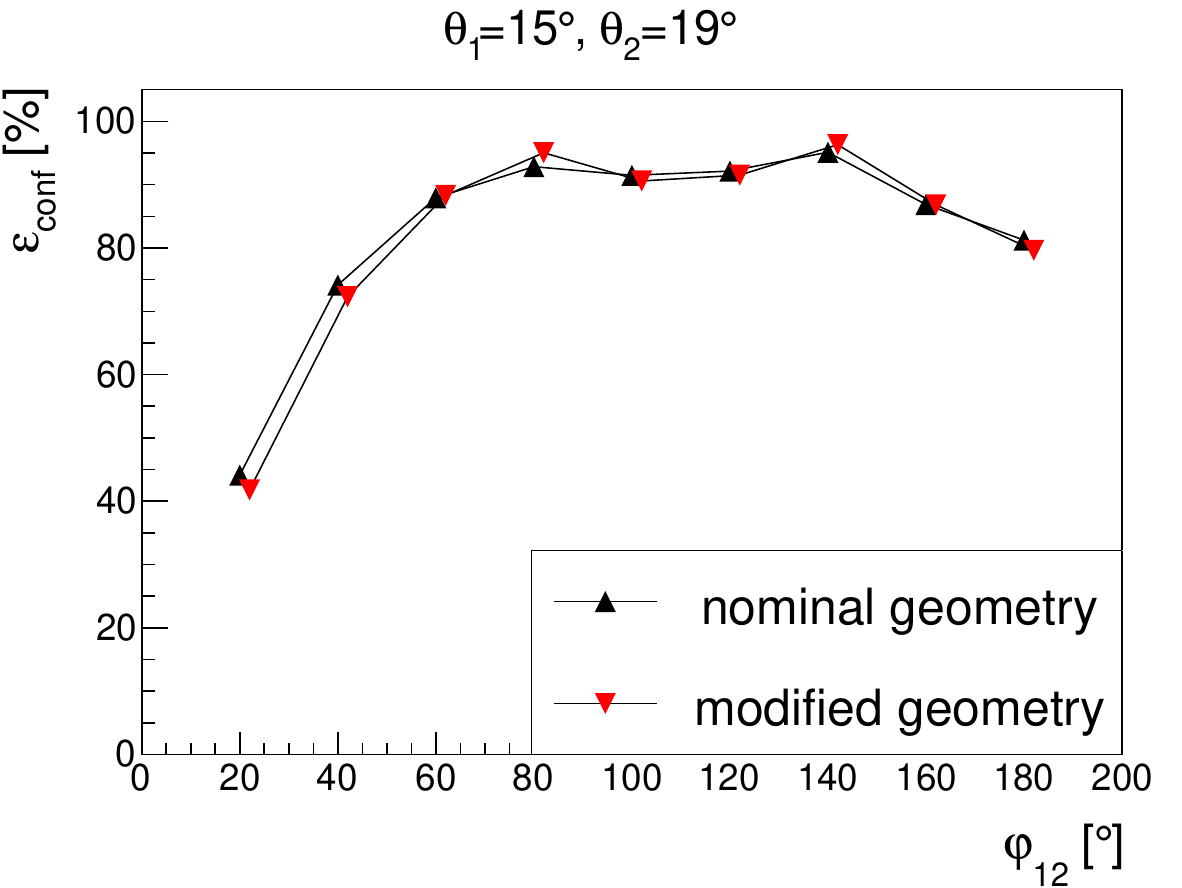}
        \caption{The configurational efficiency simulated for an nominal geometry (black triangles) and the one determined after rotating the $E$~detector (red triangles). Red triangles have been slightly shifted horizontally for better presentation of the results. 
        }
        \label{conf_eff_syst}
    \end{figure}

    \item Energy calibration \label{cal_syst}

The total impact of energy calibration, angle reconstruction, and detector efficiency correction on systematic uncertainty was estimated at 1\%. More details can be found in Ref.~\cite{Lobejko_PhD}.

    \item Trigger efficiency

The breakup events were collected with the coincidence trigger, T2 (see Table~\ref{triggers_fig}). On the other hand, we normalize breakup to the elastic events collected with a~minimum bias trigger T1. Any differences in the trigger efficiencies will not be canceled in the ratio. In order to check the relative difference of trigger efficiencies, breakup events were analyzed with a~single trigger. The number of proton pairs obtained for a~wide angular range $\theta_{1}\in(14^{\circ},26^{\circ})$, $\theta_{2}\in(14^{\circ},26^{\circ})$, and $\varphi_{12}\in(60^{\circ},120^{\circ})$ with a~single trigger has been adequately scaled ($T_1=2^4$) and compared with the number of events collected with the coincidence trigger. The accuracy of this comparison is limited by the statistics, and the estimated effect is about 3\%. Since the coincidence trigger provides a~lower number of counts, the uncertainty was implemented as an asymmetric error.

    \item Hadronic interactions

The correction of the losses due to hadronic interactions of protons and deuterons was calculated based on simulations. The estimation of these losses based on experimental data shows that both for breakup protons \cite{Parol20} and elastically scattered deuterons (this experiment, see Ref.~\cite{Lobejko_PhD}), the simulated corrections are lower than the experimental ones  by about 10-12\%, which translates to the 2\% of the corrected value. If indeed the simulation slightly underestimates the hadronic interaction losses, it happens for both protons and deuterons and the impact on the cross section result is partially canceled in the normalization and the final uncertainty was estimated at 1\%.   

\end{enumerate}

\section{Experimental results and comparison with theory} \label{results}

The differential cross section for the $^2$H(p,pp)n breakup reaction at 108~MeV was analyzed for 84 angular configurations, which corresponds to over 500 data points. The cross section distributions in function of $S$-variable are presented in Appendix A. 

The data are compared with a set of theoretical predictions based on two realistic  potentials of the  nucleon-nucleon (NN) interactions, CD-Bonn and Argonne V18. 
Calculations of Wita\l a et al.\cite{Witala1998} were performed with the CD-Bonn potential, either  alone (later referred to as CDB Witala) or augmented with the Tucsone-Melbourne 99 3NF model (CDB+TM99). Deltuva's calculations \cite{Deltuva2005} with the CD-Bonn potential (CDB Deltuva) use the approach developed by the Lisbon-Hanover-Vilnius group to  implement 3NF by adding $\Delta$-isobar excitation using the Coupled-Channel approach (CDB+$\Delta$). We also compare our data with his calculations based on the Argonne V18 potentials (AV18), and implementing the Urbana-Illinois IX 3NF model (AV18+UIX). Moreover, the set of calculations by A.~Deltuva included also the Coulomb interaction (correspondingly CDB+C, CDB+$\Delta$+C, AV18+C and AV18+UIX+C).
The two theoretical groups applied different assumptions  on isospin structure. The calculations of Wita\l a et al. were performed with $np$ interaction in the 1$s_0$ wave, while calculations of Deltuva were performed using both $pp$ and $np$ interactions in all isospin triplet waves, including 1/2 and 3/2 total 3N isospin components.

\begin{figure}[]
        \centering
        \includegraphics[width=0.4\textwidth]{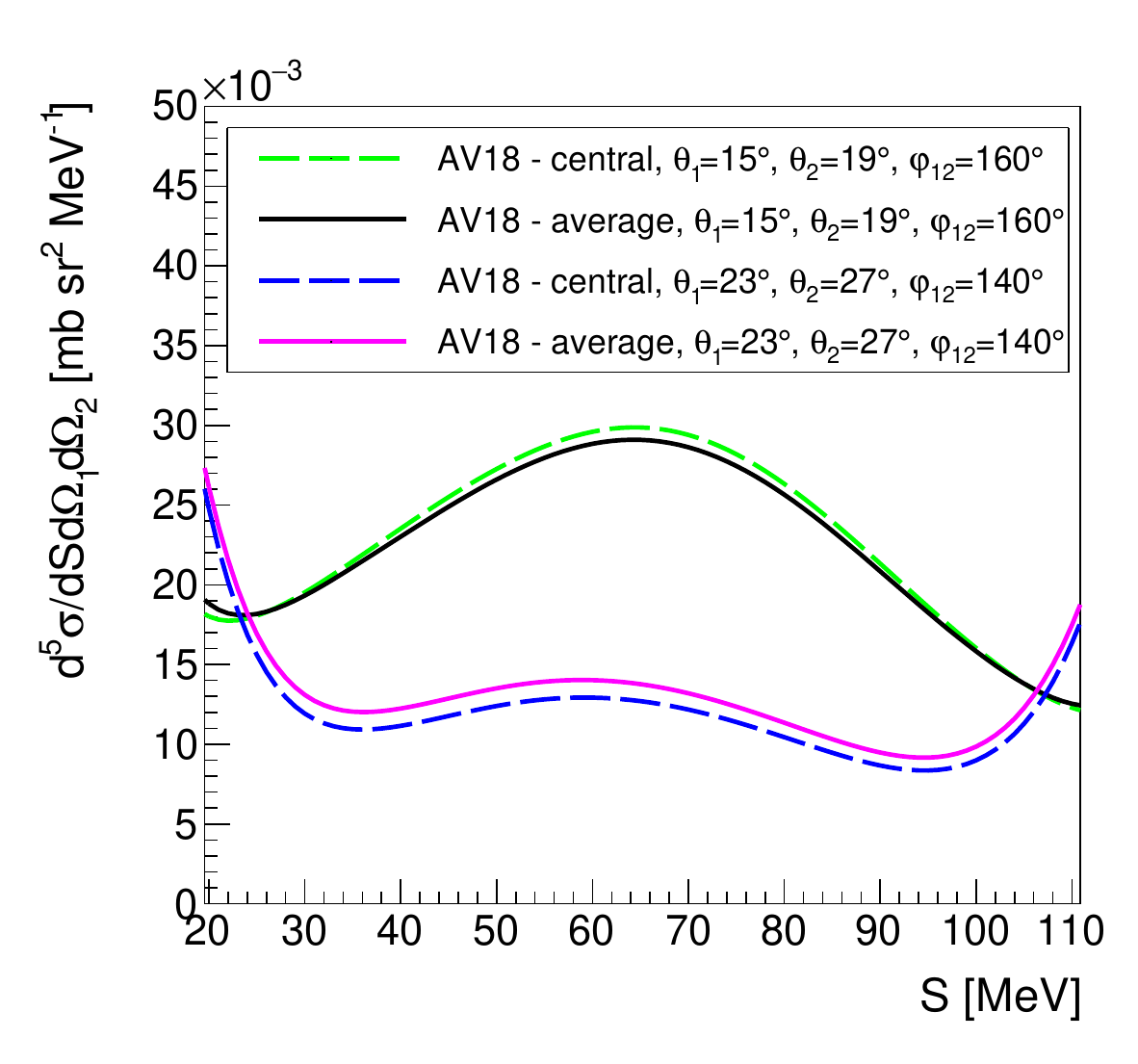}
        \caption{Theoretical calculations for the AV18 potential for two selected angular configurations: $\theta_{1}=15^{\circ}$, $\theta_{2}=19^{\circ}$, $\varphi_{12}=160^{\circ}$ and $\theta_{1}=23^{\circ}$, $\theta_{2}=27^{\circ}$, $\varphi_{12}=140^{\circ}$ (green and blue dashed lines, respectively). Black and magenta solid lines show the calculations after averaging them over acceptance ranges (see text).
       }
        \label{th_cr}
    \end{figure}

Our data analysis accepted breakup events falling into certain angular ranges  around the central configuration: 4$^{\circ}$ and 20$^{\circ}$ wide bins for the polar and azimuthal angles, respectively. Due to significant variations of the cross section within these ranges, averaging of the theoretical values over the same range was necessary. Therefore, the theoretical cross-sections have been calculated for the central configuration and all the combinations: $\theta_{1} \pm 2^{\circ}$, $\theta_{2} \pm 2^{\circ}$, and $\varphi_{12} \pm 10^{\circ}$.  Since all the calculations were non-relativistic, and a~different kinematic curve characterizes each configuration, therefore we projected all the theoretical distributions (in the $E_1$ vs $E_2$ plane) onto a~common, central relativistic kinematics. Afterwards, we determined the average of the theoretical cross section values, weighted with the solid angle of a~particular configuration for each step of the $S$-variable  and  averaged the cross section over the $S$ bin ($\pm$ 4 MeV). Finally, we obtained the averaged theoretical distribution to be compared with the experimental results. Examples of calculations with the AV18 potential for two chosen configurations is shown in Fig.~\ref{th_cr}. We present two configurations, one with a~typical and the other with a~pronounced averaging effect. The cross section distributions  obtained for central kinematics (dashed lines) are compared to the results of the above described  procedure of averaging and projecting onto relativistic kinematics (solid lines). The effect of averaging is small but not negligible.

    \begin{figure}
        \centering
        \includegraphics[width=0.4\textwidth]{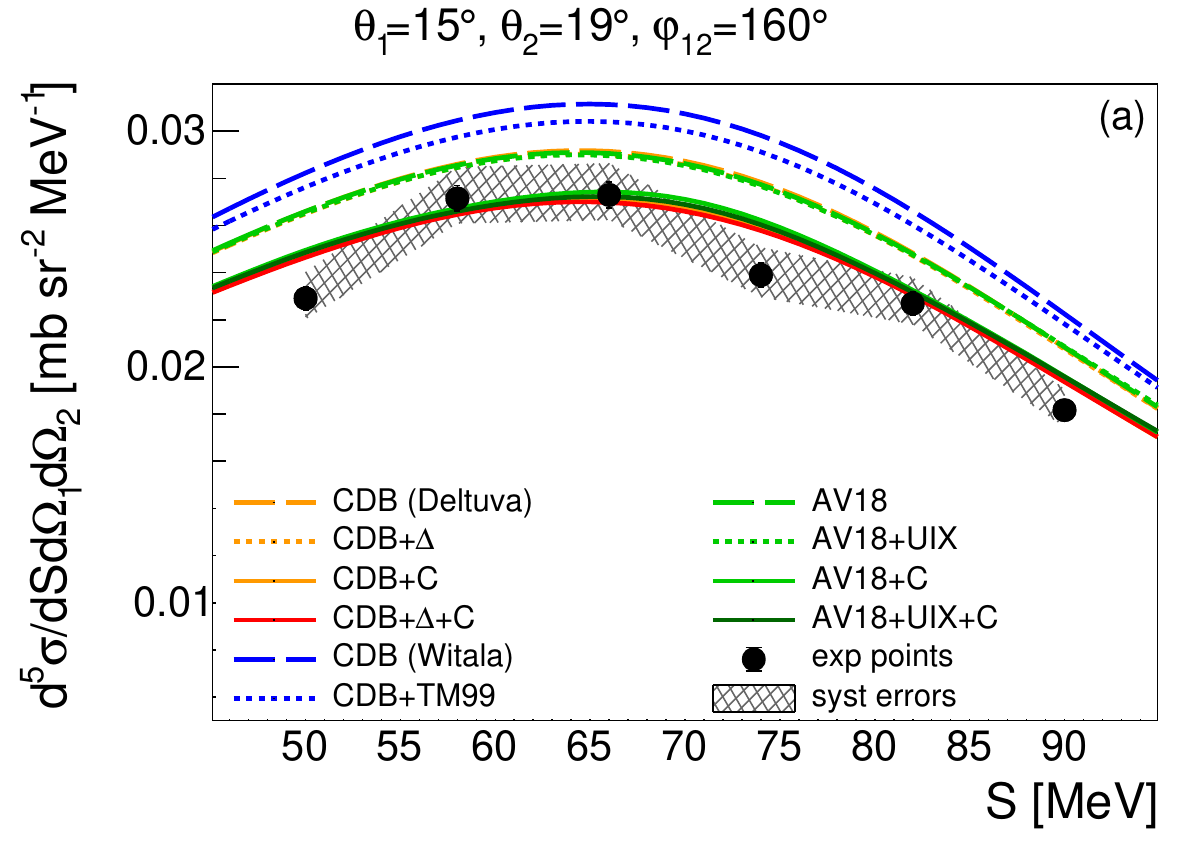}
        \includegraphics[width=0.4\textwidth]{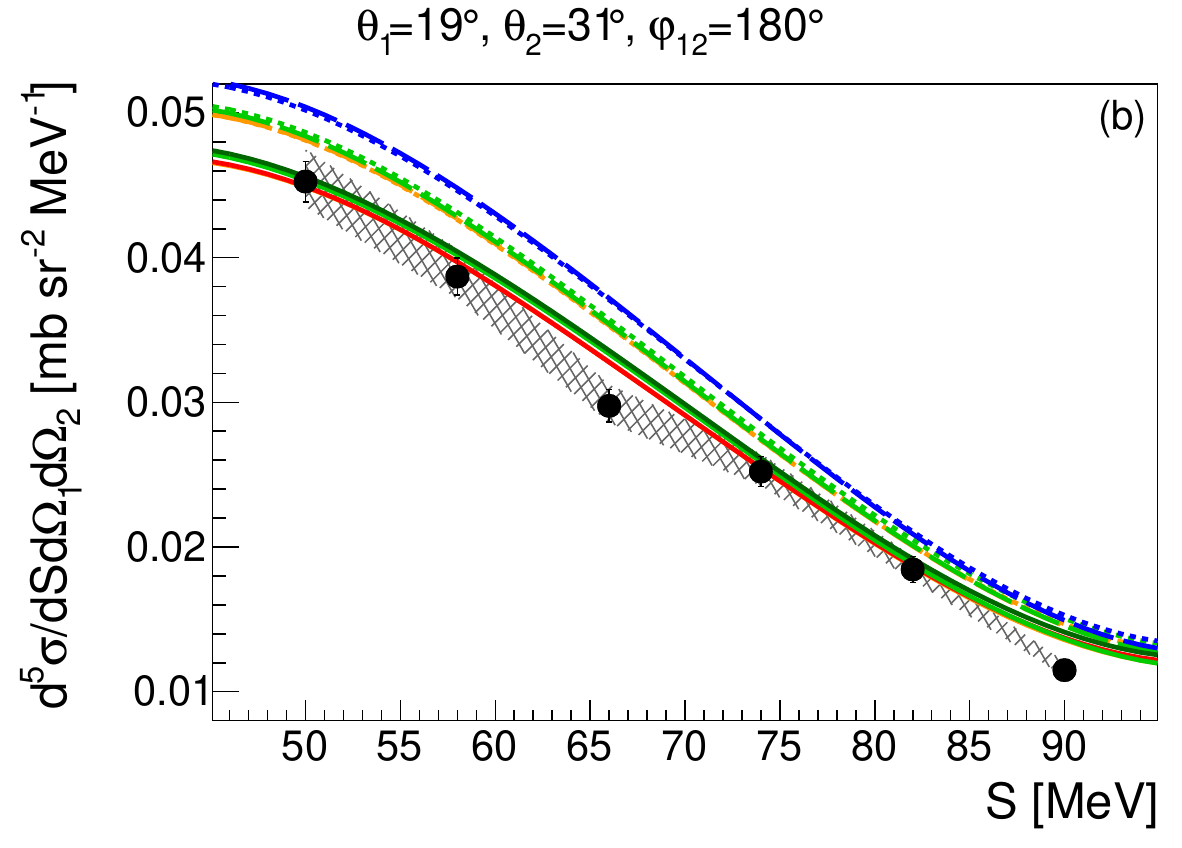}
        \includegraphics[width=0.4\textwidth]{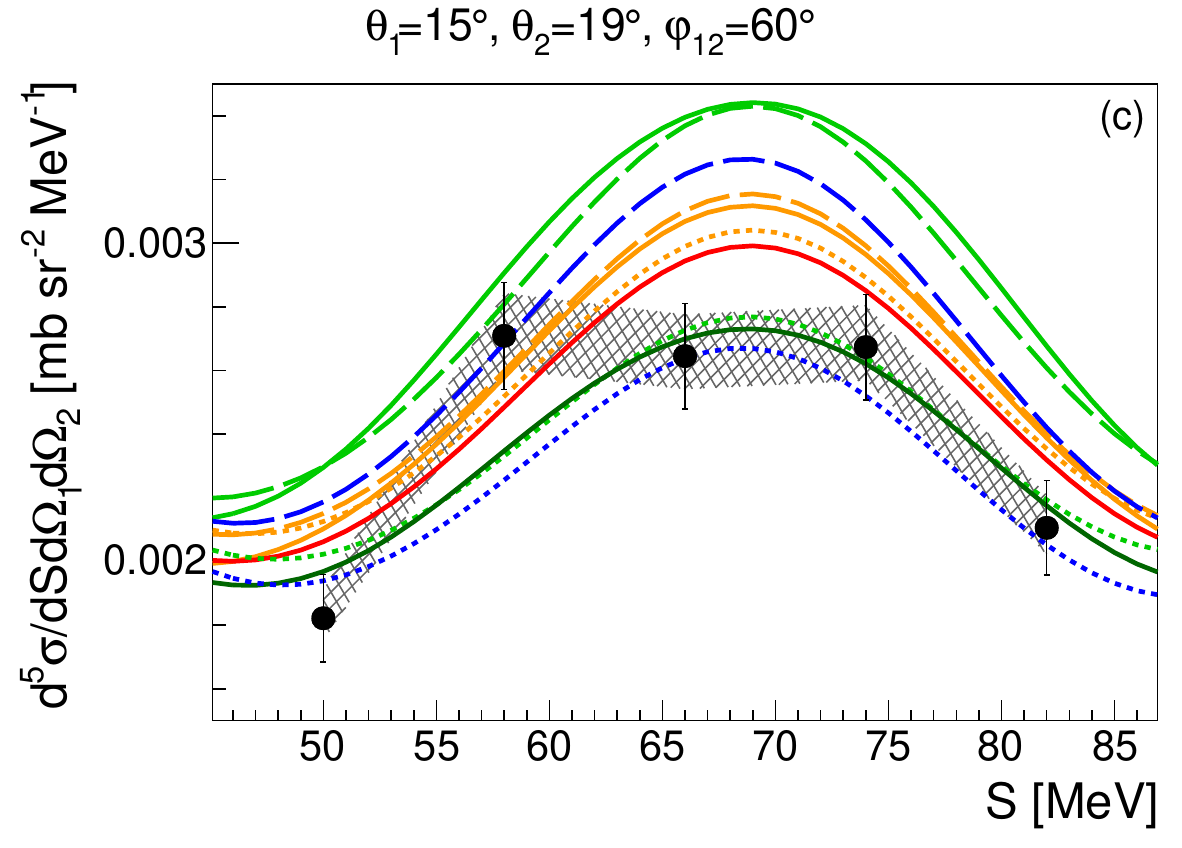}
        \includegraphics[width=0.4\textwidth]{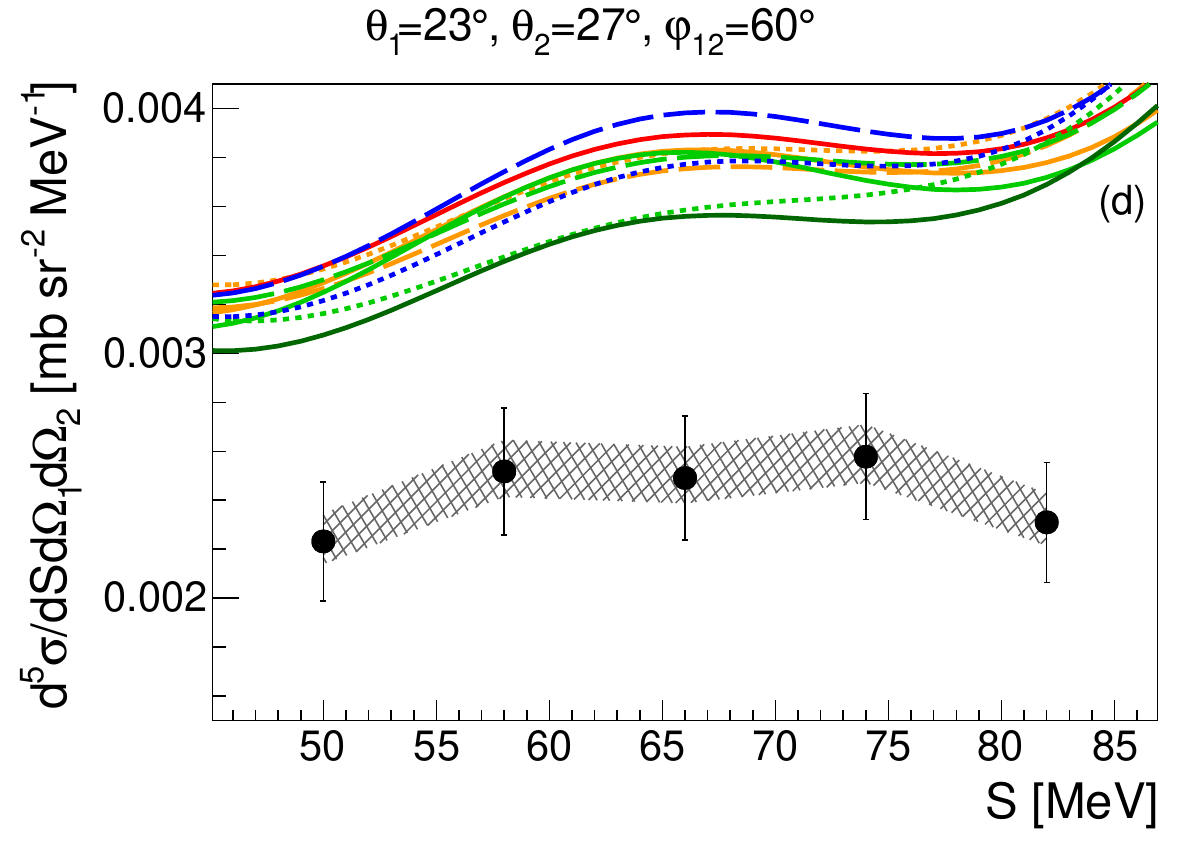}
        \caption{Examples of the differential cross section distributions for the $pd$ breakup reaction at 108 MeV.
        Black points represent experimental data with statistical errors, and gray bands illustrate their systematic uncertainties. The theoretical calculations are shown as 
        lines, see the legend in the top panel. The pure NN potentials are depicted by dashed lines, while their combinations with 3NF are shown as dotted lines. The calculations including Coulomb interaction are shown as solid lines. 
        }
        \label{br_cross_sec_rys}
    \end{figure}

Examples of the experimental differential cross section  for the $pd$ breakup reaction, compared to the theoretical calculations, are presented in Fig.~\ref{br_cross_sec_rys}. The top two panels, (a) and (b), show configurations with large $\varphi_{12}$ values. They are characterized with relatively large cross sections, thus systematic uncertainties dominate over statistical ones. The bottom two panels, (c) and (d), present configurations with $\varphi_{12}=60^{\circ}$, characterized by an order of magnitude lower cross section and thus dominating statistical errors. Generally, in regions of relatively low cross section,  we observe a significant sensitivity of the result to the details of the dynamics beyond the leading NN terms, which is reflected in a wide range of theoretical values. At the largest studied polar angles (see panel (d)), discrepancy between the data and all the theories is observed.

    \subsection{Reduced chi-squared analysis}    

A~quantitative comparison between the experimental data and theoretical predictions was performed on the basis of a~reduced chi-square, redefined as follows: 
    \begin{equation}\label{chi2_eq}
        \chi^{2}_{red} = \frac{1}{N} \sum_{i=1}^{N} \left(\frac{\sigma_{i}^{exp} - \sigma_{i}^{th}} {\Delta\sigma_{i}^{tot}}\right)^{2},
    \end{equation}
where $\sigma_{i}^{exp}$ and $\sigma_{i}^{th}$ were the experimental and theoretical values of the cross section given for each $i-th$ point, respectively, while the $\Delta\sigma_{i}^{tot}$ was the total uncertainty of each experimental point, calculated as $\Delta\sigma_{i}^{tot} = \sqrt{(\sigma_{stat}^{2} + \sigma_{syst}^{2})}$, with $\sigma_{stat}$ -  a~statistical uncertainty, and $\sigma_{syst}$ - a~systematic uncertainty. $\chi^{2}$ per degree of freedom was determined globally and for specific data subsets. 
    
     \begin{figure}
        \centering
        \includegraphics[width=0.44\textwidth]{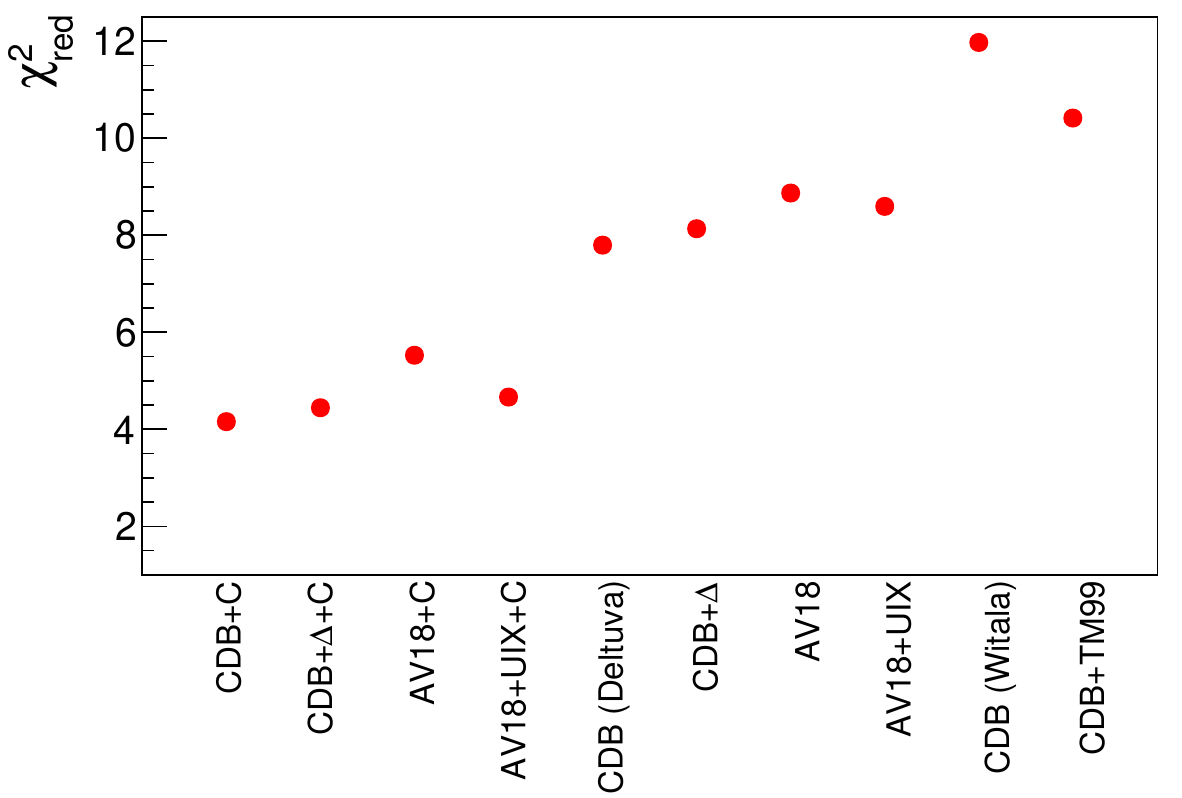}
        \caption{The global $\chi^{2}_{red}$ derived for all data points, individually for each theory.}
        \label{chi2_gl}
    \end{figure}

At the outset $\chi^{2}_{red}$ has been computed as a~global factor, where $N$ corresponds to the total number of data points. 
In Fig.~\ref{chi2_gl} we observe clear dependence of the data description on the theoretical model, which is reflected in  a~global $\chi^{2}_{red}$ varying from approximately 4~to 12. 
The highest values correspond to calculations performed by Witala et al. Admittedly, adding the~TM99 three-nucleon force improves the $\chi^{2}_{red}$ compared to the pure CD-Bonn potential, but the discrepancy remains. There are also four results grouped around $\chi^{2}_{red} \approx 8$ corresponding of the theoretical calculations performed by Deltuva. Here, the effect of adding the $\Delta$-isobar to the CD-Bonn potential and the Urbana IX 3NF to the AV18 potential is very small, respectively slightly increasing or decreasing the global $\chi^{2}_{red}$.
The global $\chi^{2}_{red}$ analysis clearly shows a~significant improvement in the description of the cross section data when the Coulomb force is taken into account. Not only the $\chi^{2}_{red}$ is reduced by a factor of two, but also the effect of Urbana IX 3NF is revealed. However, the size of this effect is similar to the difference in description given by  calculations with two different NN potentials.

\begin{figure}[]
        \centering
        \includegraphics[width=0.23\textwidth]{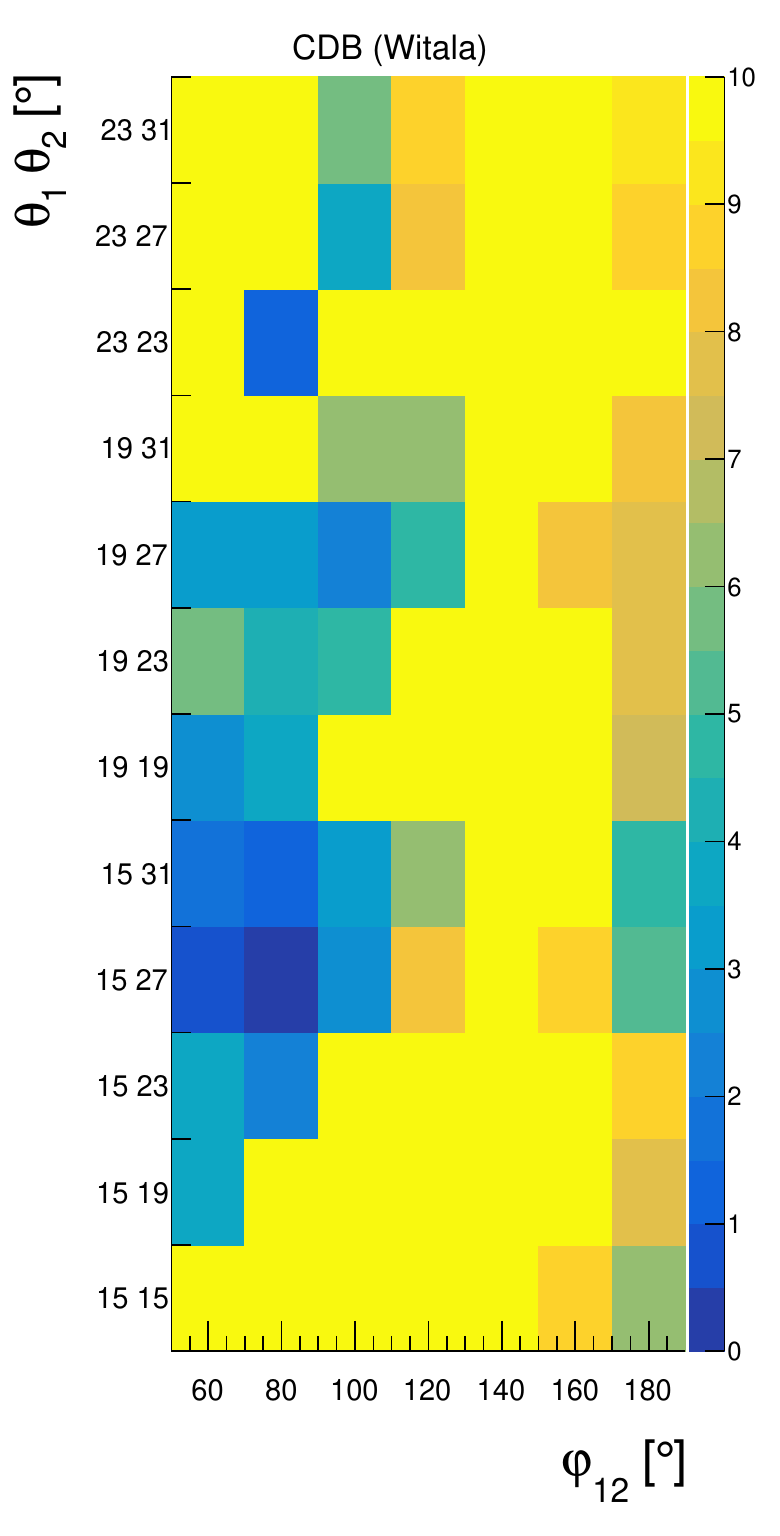}
        \includegraphics[width=0.23\textwidth]{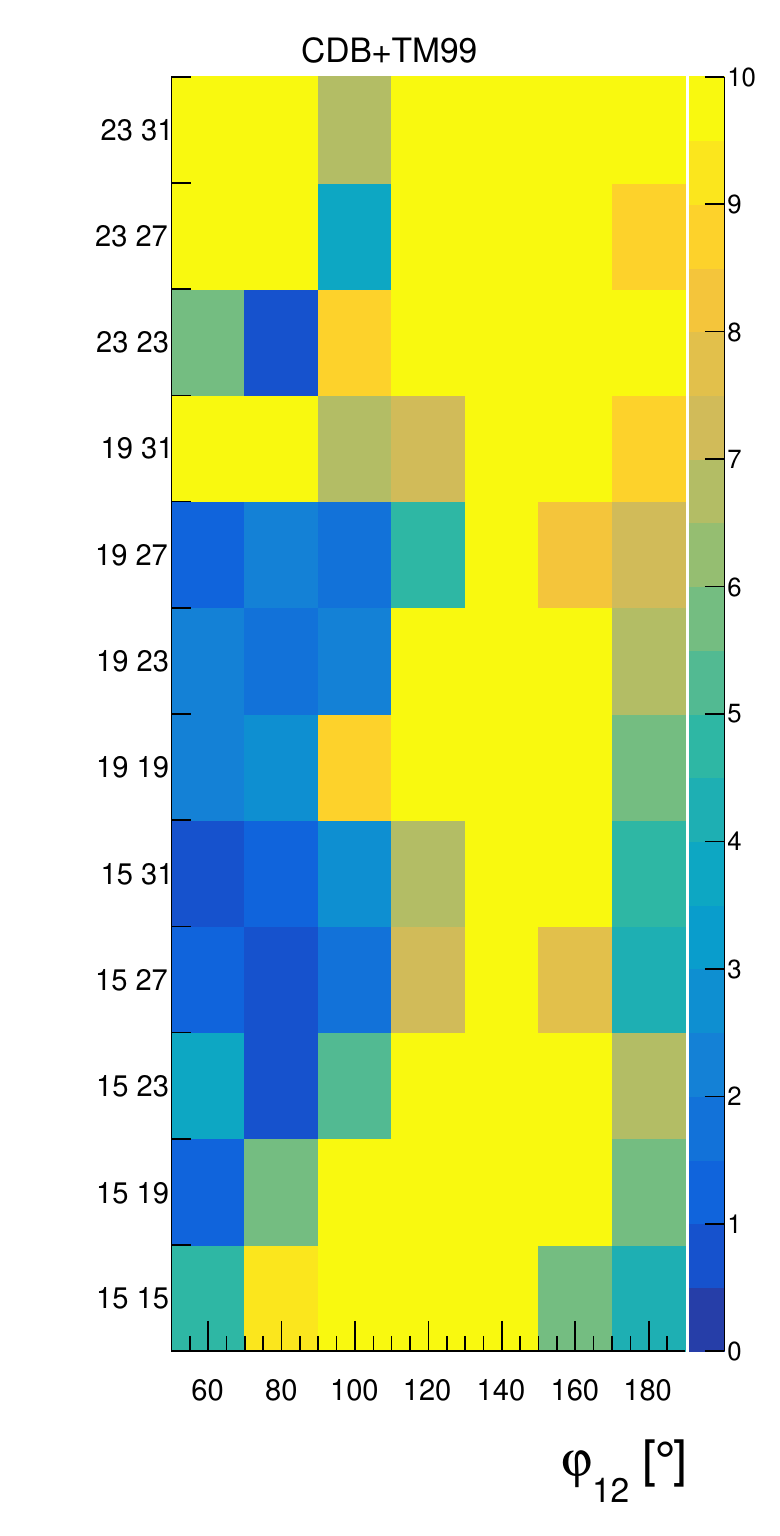}
        \caption{The $\chi^{2}_{red}$ results obtained individually for each angular configurations, presented as a 2-dimensional map defined by combination of polar and relative azimuthal angles.  The data are compared with two selected theories: CDB (Witala) and CDB+TM99.}
        \label{chi2_rys1}
    \end{figure}

\begin{figure*}[]
        \centering
        \includegraphics[width=0.23\textwidth]{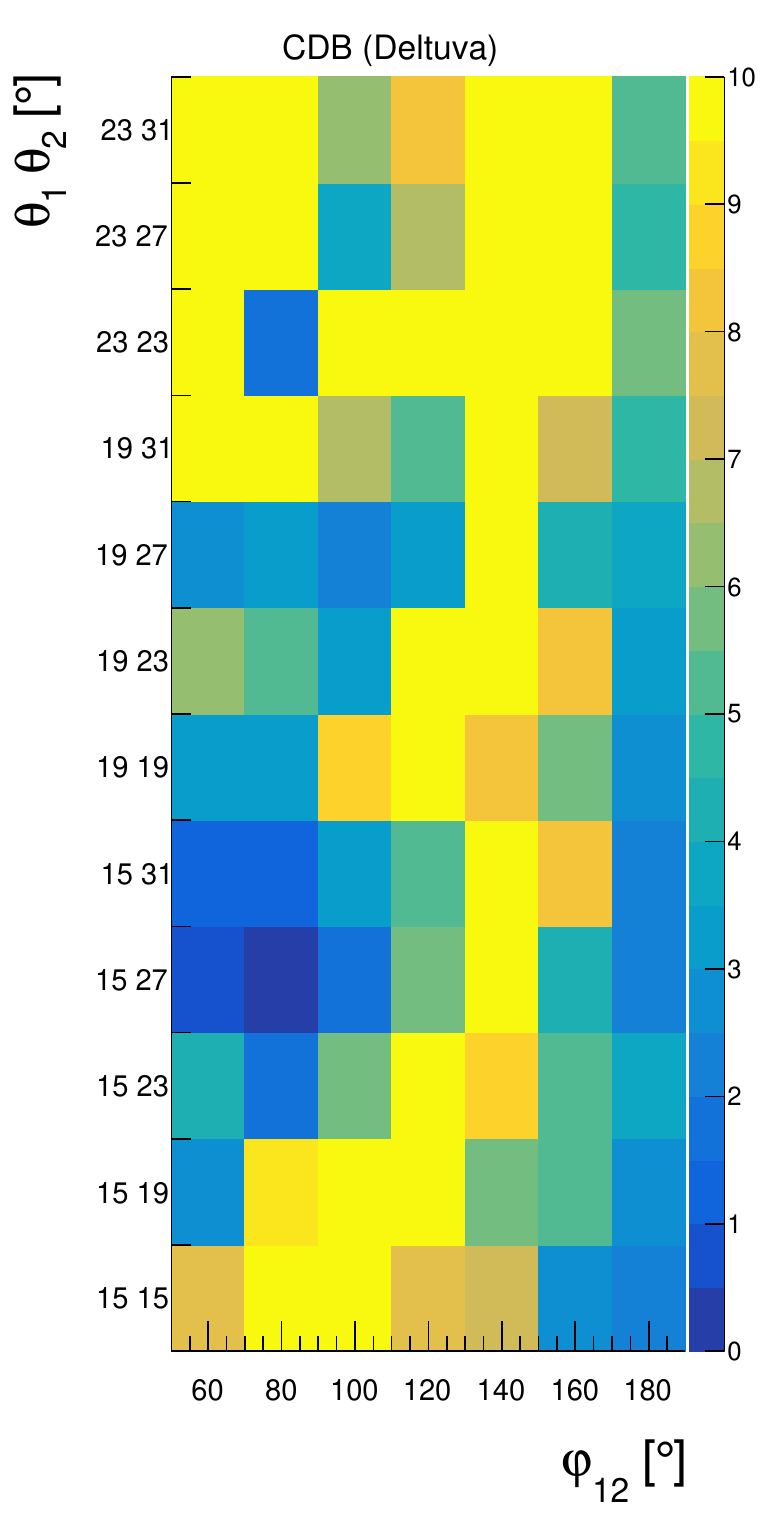}
        \includegraphics[width=0.23\textwidth]{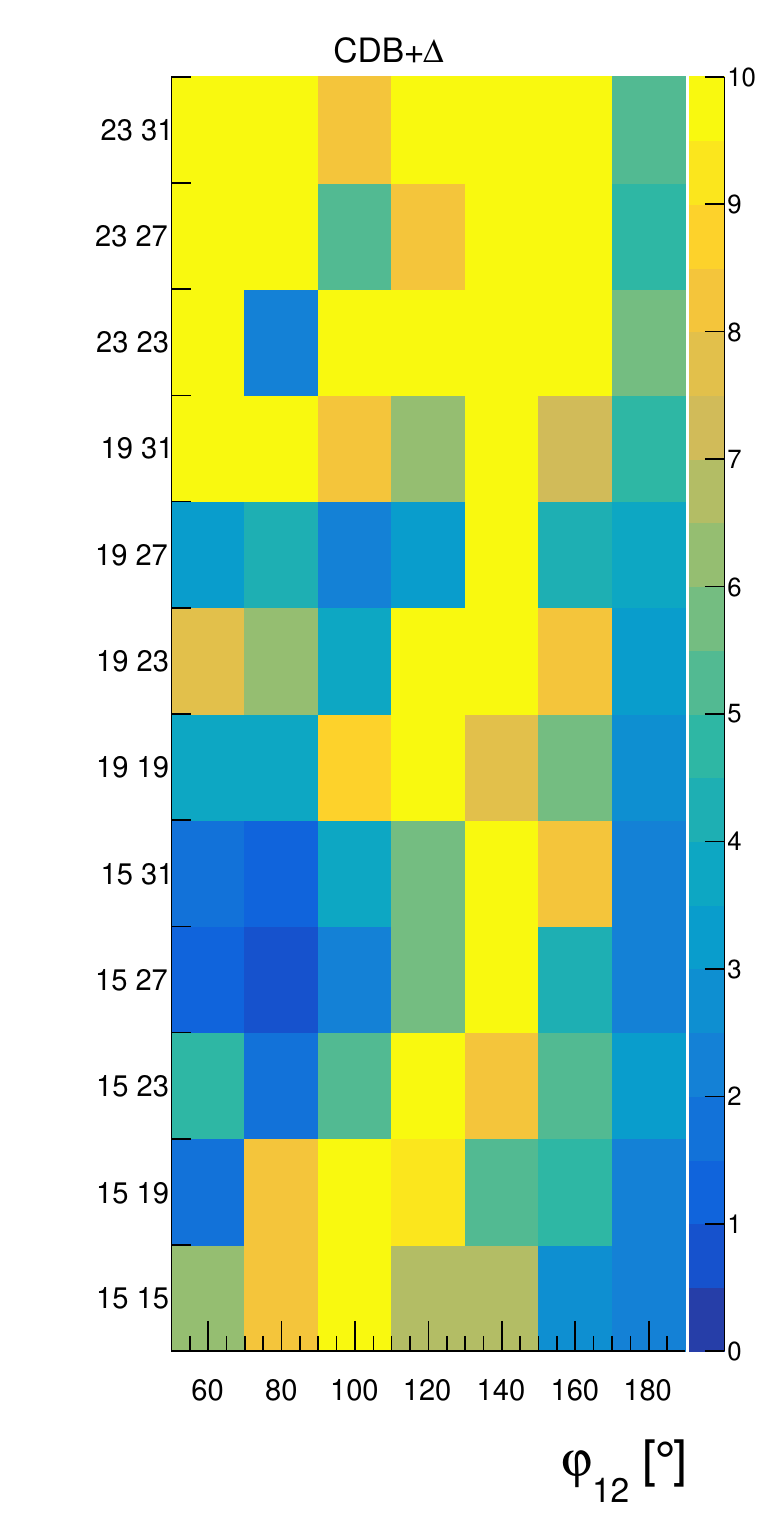}
        \includegraphics[width=0.23\textwidth]{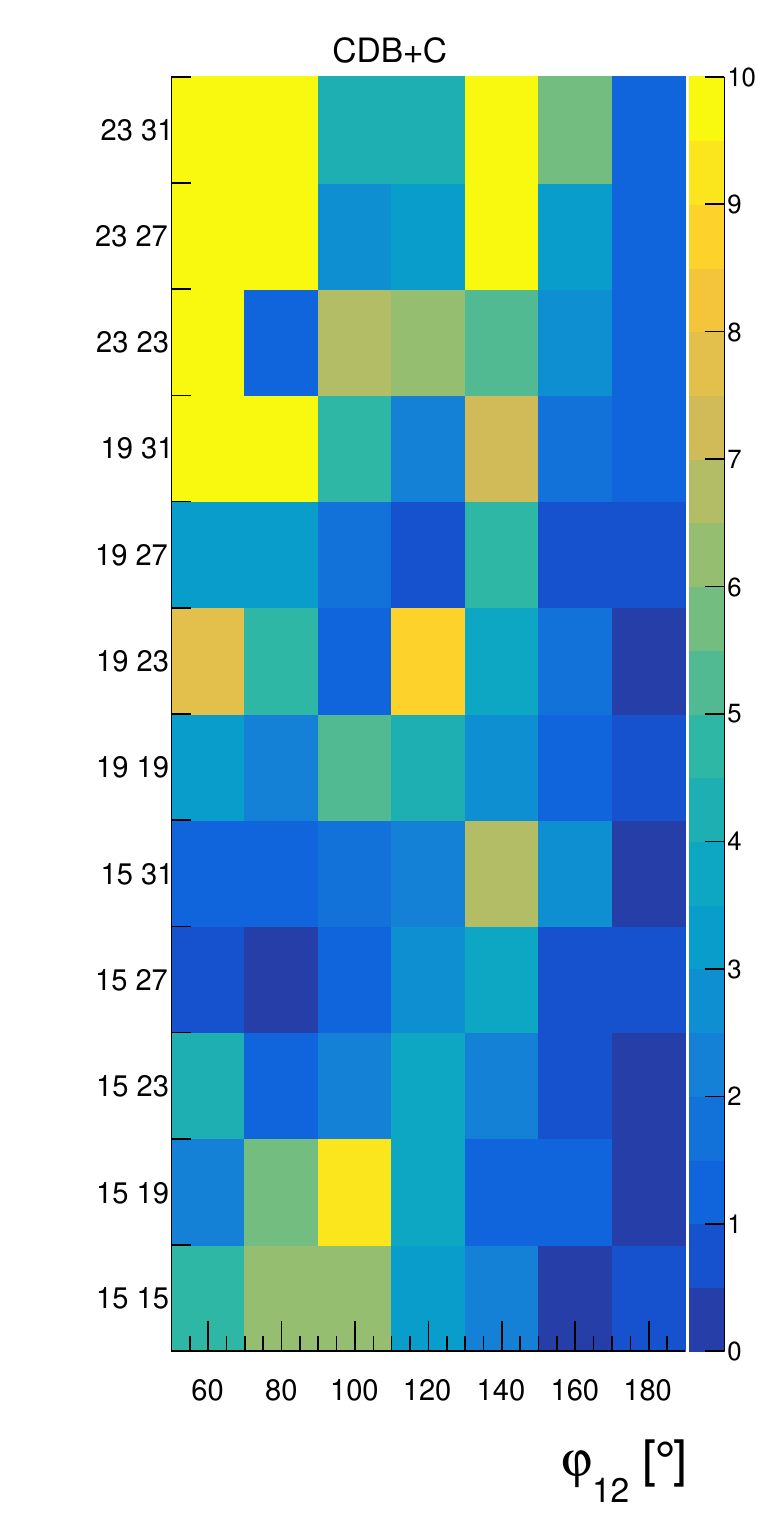}
        \includegraphics[width=0.23\textwidth]{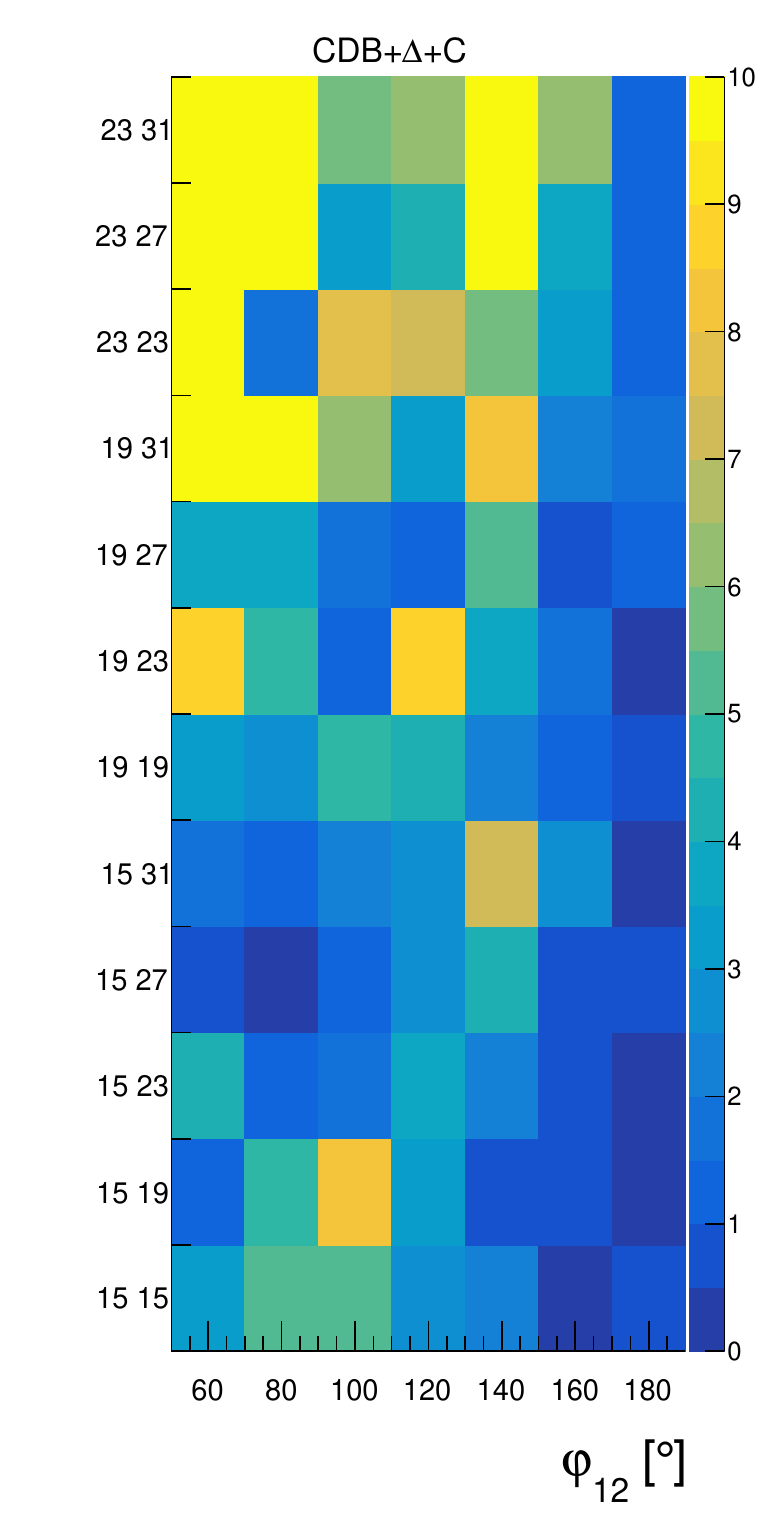}
        \caption{Similar as in Fig.~\ref{chi2_rys1}, but for different set of theoretical calculations: CDB (Deltuva), CDB+$\Delta$, CDB+C, CDB+$\Delta$+C.}
        \label{chi2_rys2}
    \end{figure*}

\begin{figure*}[]
        \centering
        \includegraphics[width=0.23\textwidth]{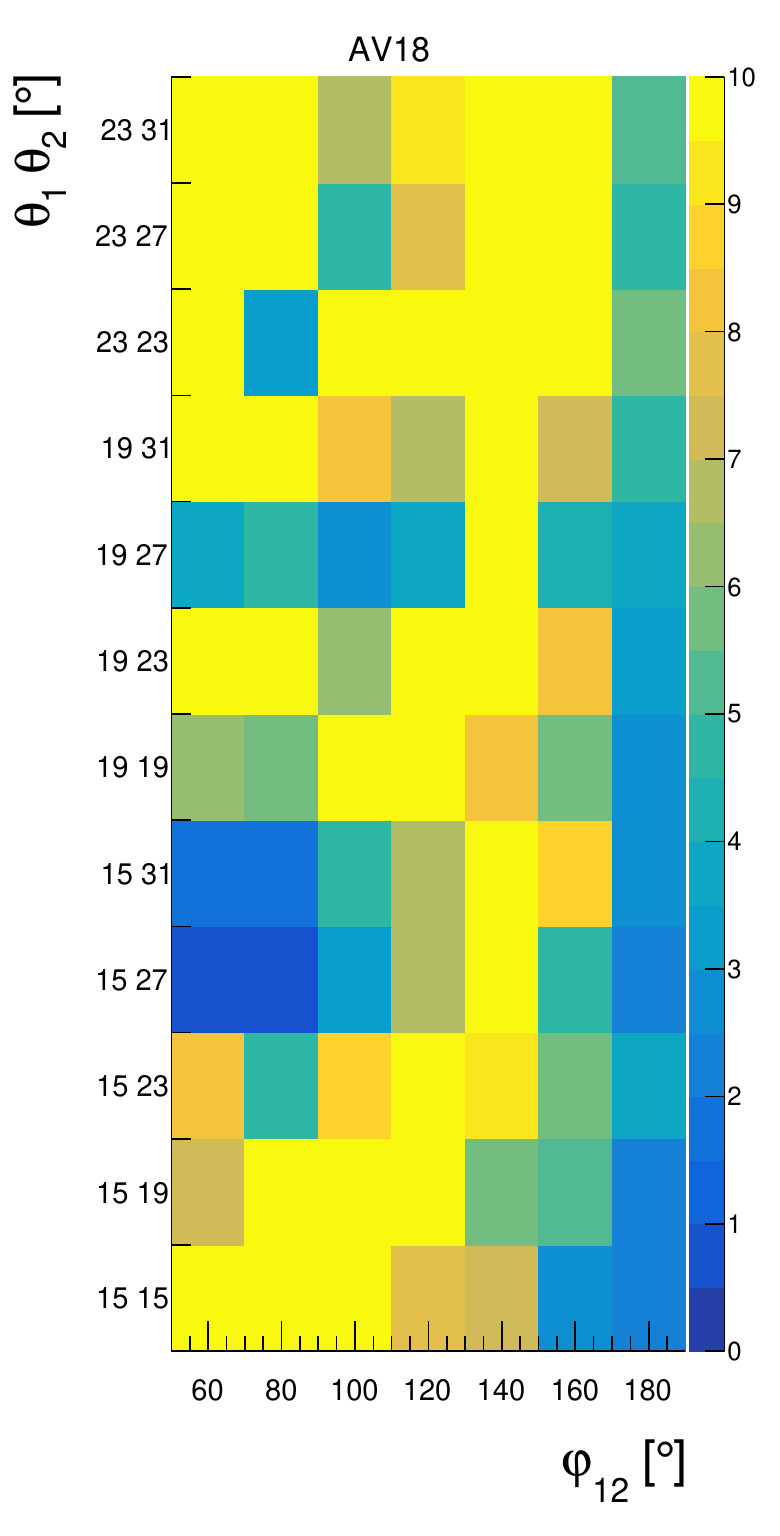}
        \includegraphics[width=0.23\textwidth]{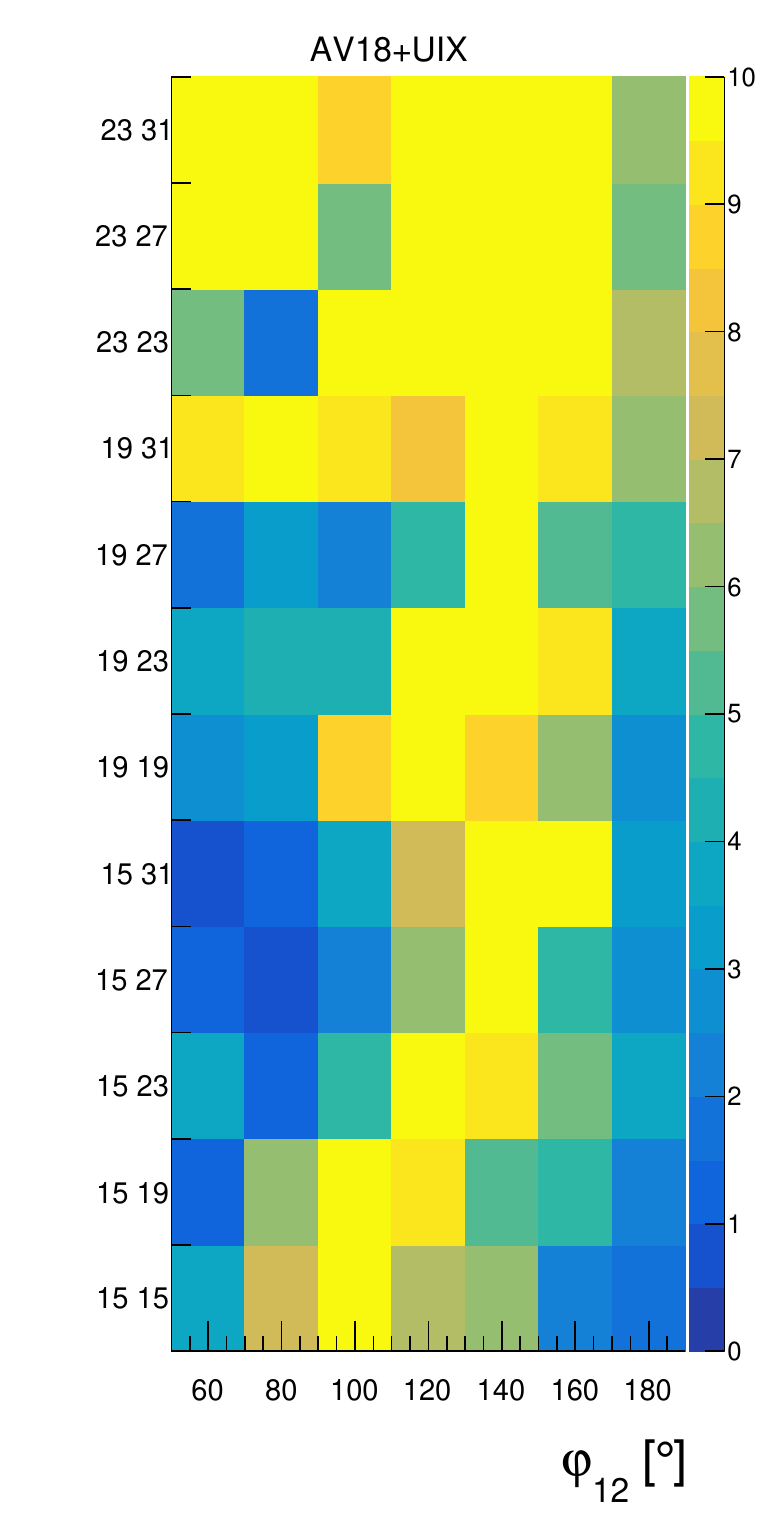}
        \includegraphics[width=0.23\textwidth]{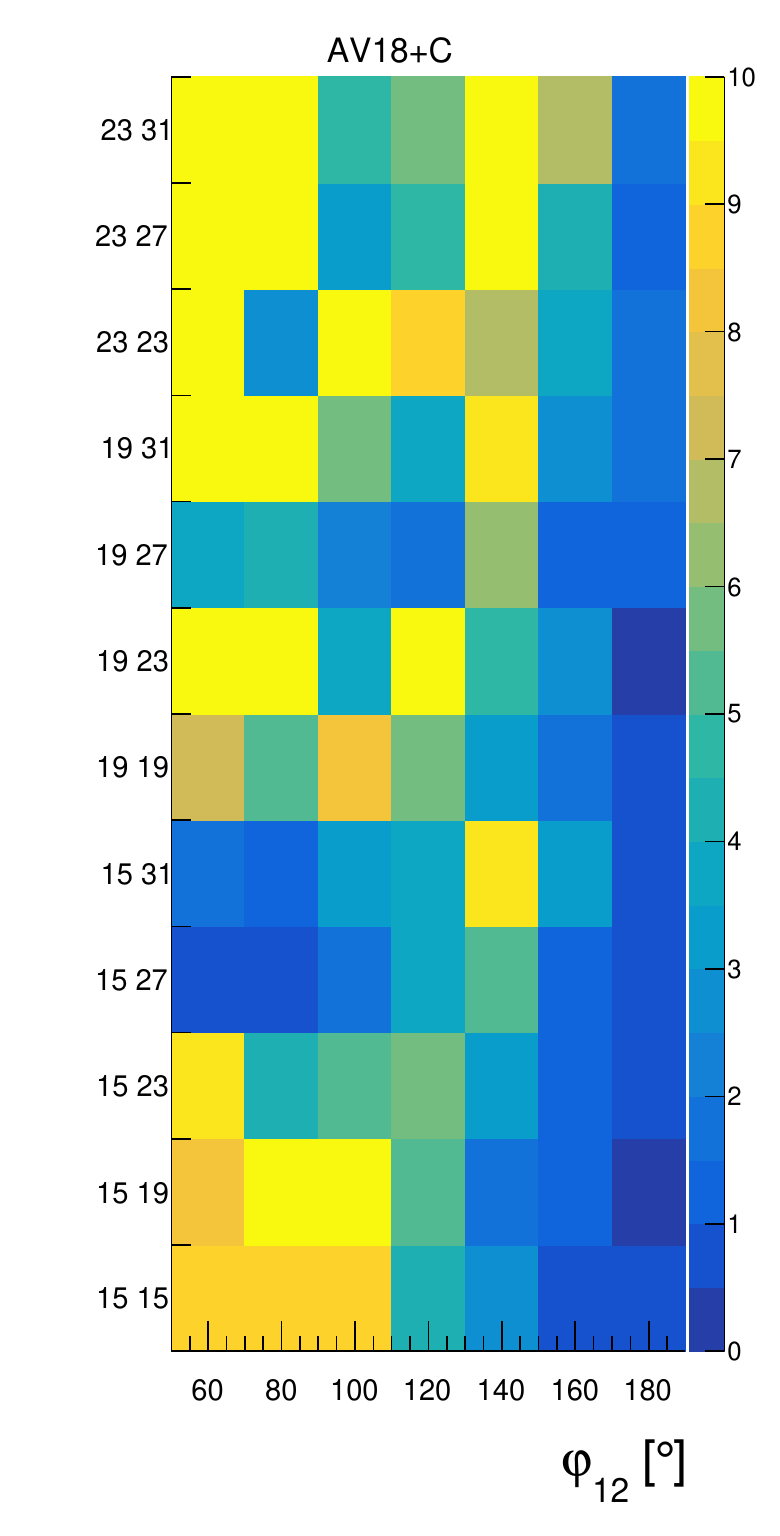}
        \includegraphics[width=0.23\textwidth]{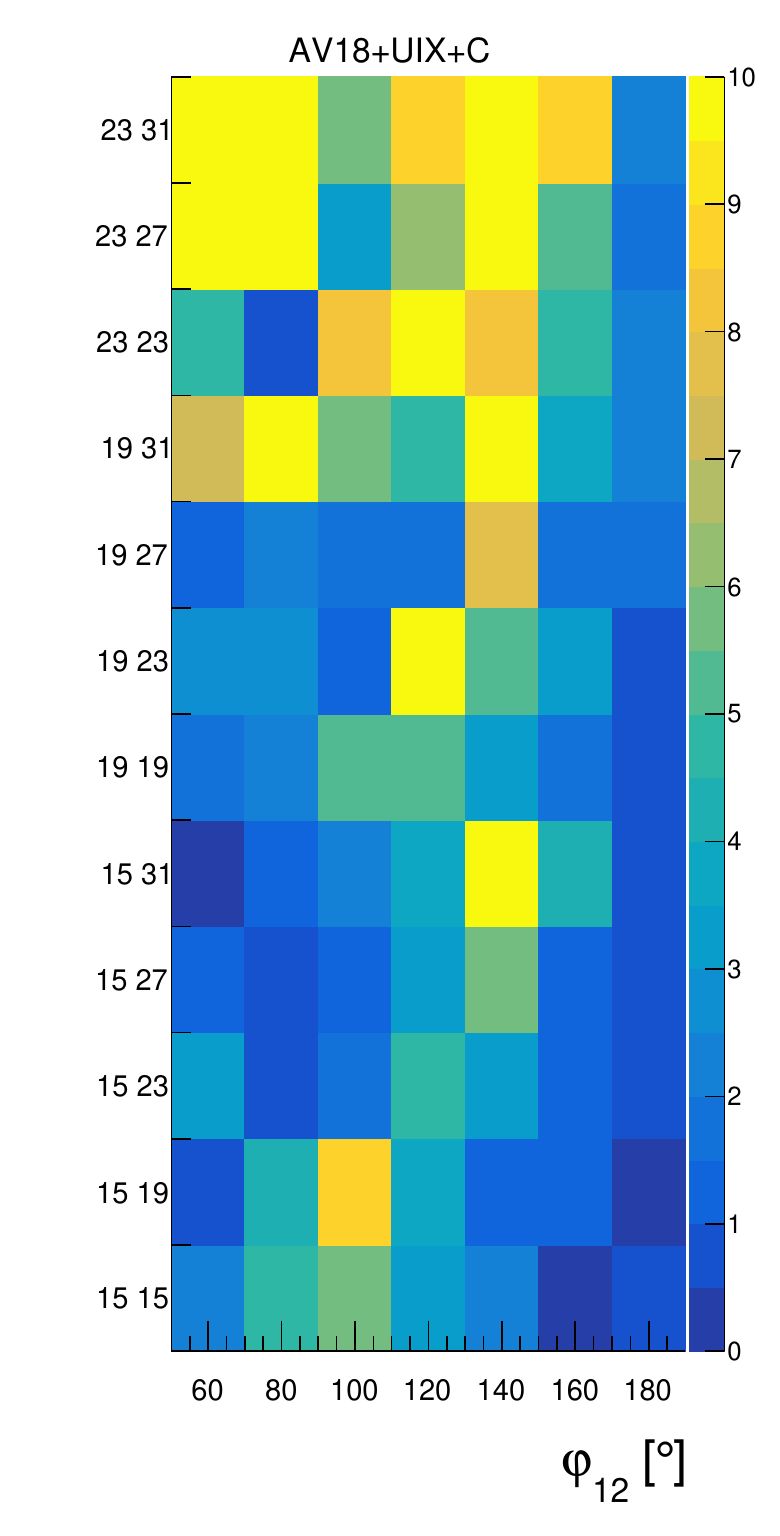}
        \caption{Similar as in Fig.~\ref{chi2_rys1}, but for different set of theoretical calculations: AV18, AV18+UIX, AV18+C, AV18+UIX+C.}
        \label{chi2_rys3}
   \end{figure*}

In the next step, the quality of the description was checked locally. $\chi^{2}_{red}$ was calculated individually for each angular configuration, where $N$~is the number of experimental points in a~given configuration. The results are presented as 2-dimensional maps, plotted for each theoretical calculation separately, see Figs.~\ref{chi2_rys1}-\ref{chi2_rys3}.

The $\chi^{2}_{red}$ maps obtained for theories without Coulomb (Fig.~\ref{chi2_rys1}, and the two left panels in Figs.~\ref{chi2_rys2} and \ref{chi2_rys3}) demonstrate deficiency of the data description in particular regions. These areas are mainly located at intermediate $\varphi_{12}$ and also in the upper part corresponding to the higher polar angles $\theta_{1}$ and $\theta_{2}$. 

\begin{figure}[]
        \centering
        \includegraphics[width=0.44\textwidth]{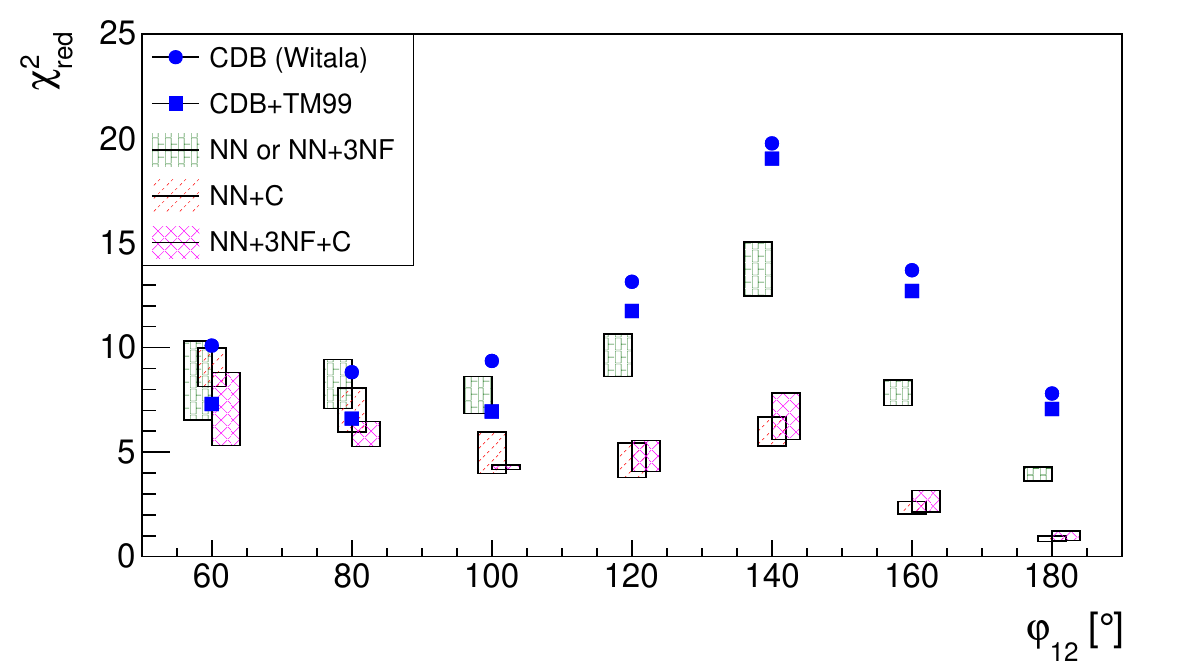}
        \includegraphics[width=0.44\textwidth]{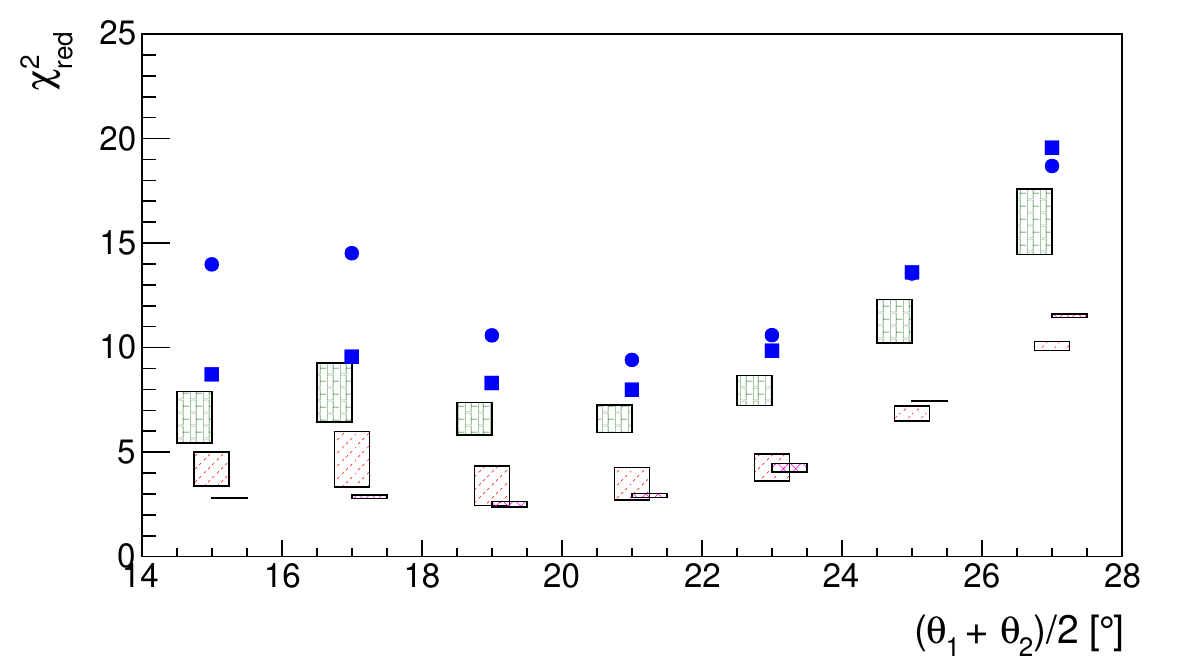}
        \caption{The $\chi^{2}_{red}$ results presented as a~function of the relative azimuthal angle $\Delta\varphi_{12}$ (\textit{upper panel}), and with respect to the average of $\theta_1$ and $\theta_2$ (\textit{lower panel}). The legend is common for both panels.}
        \label{chi2_phi}
    \end{figure}

The maps enable us to observe the quality of the theoretical description for each of the configurations separately, but for better visibility of the effects they have been truncated to a maximal value of 10. The full range of $\chi^{2}_{red}$ can be observed separately for the relative azimuthal angles $\varphi_{12}$ (Fig.~\ref{chi2_phi}, \textit{upper panel}) and polar angles $\theta_{1}$ and $\theta_{2}$ (Fig.~\ref{chi2_phi}, \textit{lower panel}). For calculations without Coulomb (blue points and squares, green shaded bars), the greatest discrepancy is for $\varphi_{12}=140^{\circ}$ and maximal polar angles.

Adding the Coulomb interaction substantially improves the situation, as shown in two right panels in Figs.~\ref{chi2_rys2} and \ref{chi2_rys3}. Additionally, we observe that all the calculations including Coulomb interaction  (red and magenta shaded bars) reach minimum of $\chi^{2}_{red}$ in a~similar region ($\varphi_{12}=100^{\circ}$, $160^{\circ}$ and $180^{\circ}$, see Fig.~\ref{chi2_phi}, \textit{upper panel}). \textit{Lower panel} in Fig.~\ref{chi2_phi} shows a~minimum of $\chi^{2}_{red}$ between $(\theta_1 + \theta_2) / 2 = 15^{\circ}$ and $(\theta_1 + \theta_2) / 2 = 21^{\circ}$, observed for all the theoretical calculations including Coulomb interaction.  

Even calculations neglecting Coulomb indicate certain improvement of description when TM99 or Urbana IX 3NF are added, in particular at the bottom-left corners of the maps in Figs.~\ref{chi2_rys1}-\ref{chi2_rys3}. This region corresponds to the lowest $\varphi_{12}$ and $\theta$'s. Once the Coulomb interaction is introduced, the possible effects of 3NF can be reliably investigated in the whole studied phase space. The overall picture is not so clear, but locally the improvement can be seen, especially at low $\varphi_{12}$, see Fig.~\ref{chi2_phi} \textit{upper panel} and an example in Fig.~\ref{br_cross_sec_rys}(c). It is also visible in the maps, in their bottom-left corners, where low $\varphi_{12}$ is combined with the lowest polar angles. This confirms the observation for theories without Coulomb. 

The remaining discrepancy between data and calculations is visible at low $\varphi_{12}$ and large $\theta$'s, i.e. in left-top corner of the maps, see also the example in Fig.~\ref{br_cross_sec_rys}(d).

\section{Summary}
The differential cross sections for the $^2$H(p,pp)n reaction at beam energy of 108 MeV have been measured with the forward Wall of the BINA detector. The data, collected  in the range of polar angles of outgoing protons from 13 to 33 degrees, were sorted into 84 configurations, corresponding to various combinations of polar and relative azimuthal angles. Taking into account binning in the energy variable $S$, over 500 data points were obtained.

The experimental results have been compared to the state-of-the-art theoretical calculations including contributions from two-nucleon interaction combined or not with the 3NF or Coulomb force. The quality of the data description was studied by performing a~reduced chi-square ($\chi^2_{red}$) analysis for each of the available models. 

The investigated region of the phase space is characterized by significant variation in the differential cross section values. Regions with the lower cross-section reveal greater sensitivity to the dynamic effects. In almost the entire studied phase space, the Coulomb effects are present. Among all the compared theoretical models, the best agreement is obtained for the calculations including Coulomb interactions (CDB+C, CDB+$\Delta$+C AV18+C, AV18+UIX+C). 
The results show importance of the Coulomb effects at the studied energy and lead to the conclusion that the Coulomb interaction has to be necessarily included into the theoretical description. It is important also in the context of future comparisons of the data with the intensively developed calculations within the Chiral Effective Field Theory.

In the studied~part of the breakup phase space, the effects of three-nucleon force are moderate or negligible. For the most sensitive configurations, we observe a larger influence of the TM99 and UrbanaIX forces than of the $\Delta$-isobar. There is also a small kinematical region where all the theories underestimate the data. There, the 3N forces (TM99 and UIX) lower the cross section values, thus improve the description, but by far not enough to eliminate the discrepancy. 

In the future, the results presented here will be extended by analyzing data from the subsequent experimental run covering a~wider angular range. Configurations close to the so-called neutron-proton final state interaction are kinematically similar to the elastic scattering and can reveal stronger sensitivity to 3NF.

\begin{acknowledgments}
This work was supported by the Polish National Science Centre under Grants\- No. 2020/37/N/ST2/02360, 2016/23/D/ST2/01703 and 2012/05/B/ST2/02556. This research was also supported in part by the Excellence Initiative – Research University Program at the Jagiellonian University in Kraków. The numerical calculations were performed on the supercomputer cluster of the JSC, Jülich, Germany.

We wish to express our gratitude to the PROTEUS cyclotron crew at the CCB for providing us with an excellent and stable beam during the experiment.

\end{acknowledgments}

\appendix

\section{}

Appendix A~contains all the differential cross section distributions 
for the $^2$H(p,pp)n breakup reaction at the beam energy of 108~MeV obtained in the analysis described in this paper. 
Each figure presents 7 configurations corresponding to individual azimuthal angles $\varphi_{12}$ from $60^{\circ}$ to $180^{\circ}$ and one combination of $\theta_{1}$ and $\theta_{2}$ specified in the caption. The legend is presented in panel 8th.

    \begin{figure*}[]
        \centering
        \includegraphics[width=1.\textwidth]{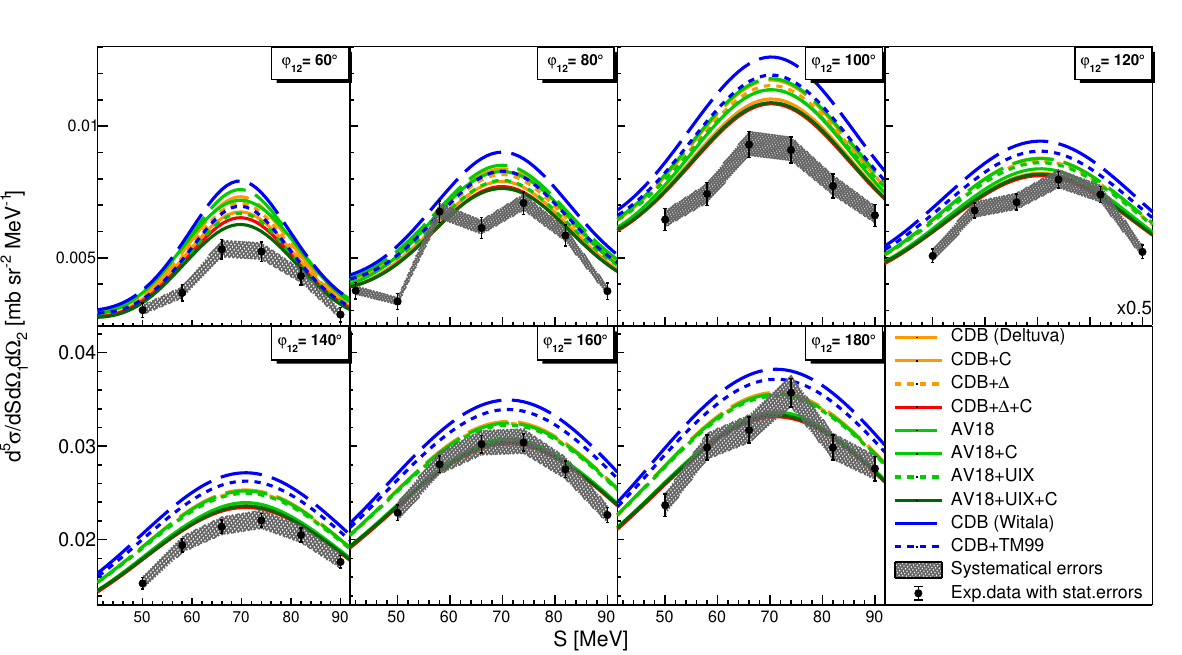}
        \caption{The differential cross section for $pd$ breakup reaction obtained for polar angles $\theta_{1}=15^{\circ}$ and $\theta_{2}=15^{\circ}$ and for a~set of azimuthal angles $\varphi_{12}$ specified in the panels. Black points represent the experimental data with statistical errors, and gray bands illustrate the systematic uncertainties. The available theoretical calculations are shown as  lines, listed in the legend. In one panel the results are scaled by 0.5 to fit the common vertical axis, as indicated in the panel.}
        \label{br1_ap}
    \end{figure*}
    
    \begin{figure*}[]
        \centering
        \includegraphics[width=1.\textwidth]{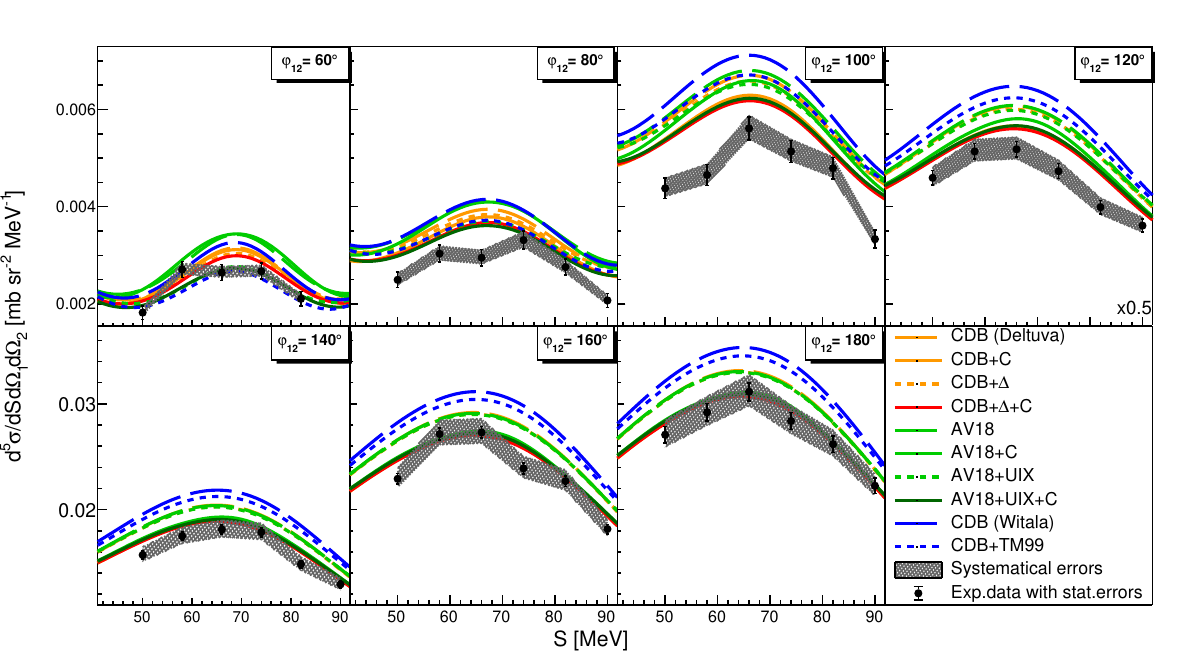}
        \caption{The same as Fig.~\ref{br1_ap} but for polar angles $\theta_{1}=15^{\circ}$ and $\theta_{2}=19^{\circ}$.}
        \label{br2_ap}
    \end{figure*}

    \begin{figure*}[]
        \centering
        \includegraphics[width=1.\textwidth]{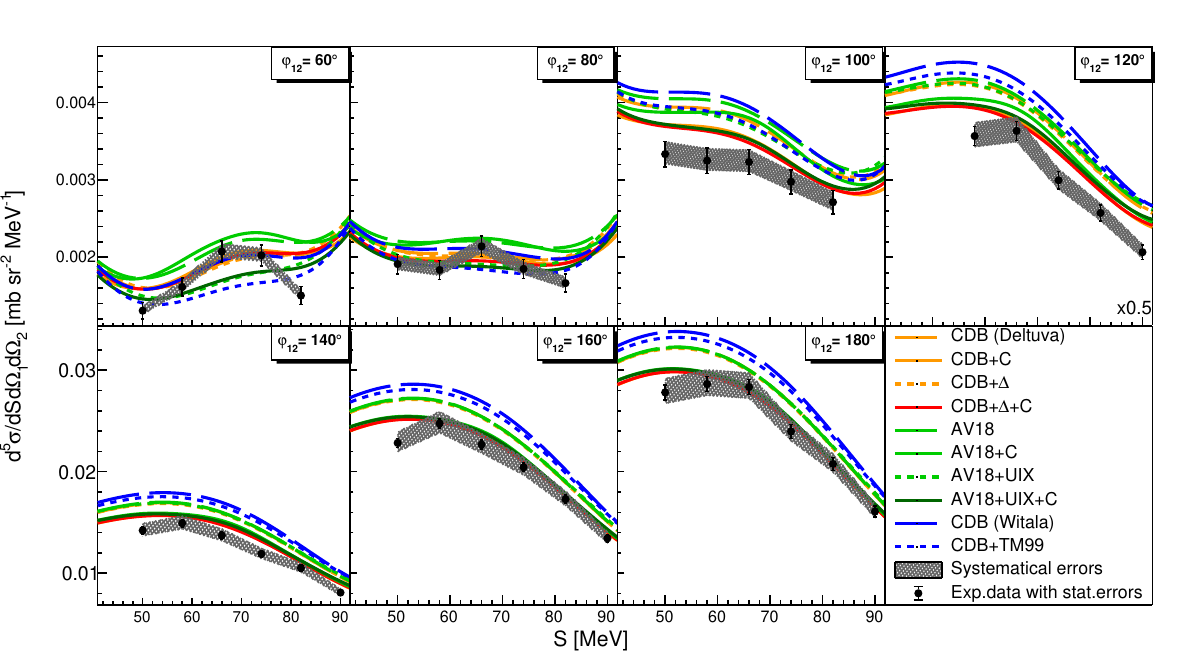}
        \caption{The same as Fig.~\ref{br1_ap} but for polar angles $\theta_{1}=15^{\circ}$ and $\theta_{2}=23^{\circ}$.}
        \label{br3_ap}
    \end{figure*}

    \begin{figure*}[]
        \centering
        \includegraphics[width=1.\textwidth]{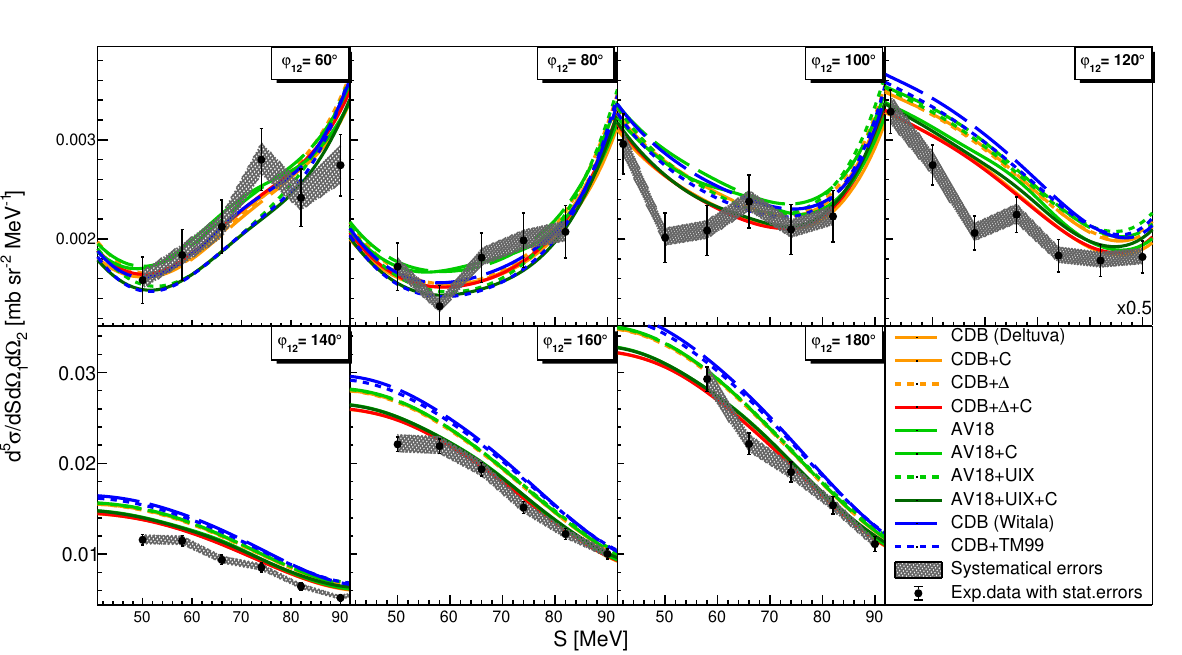}
        \caption{The same as Fig.~\ref{br1_ap} but for polar angles $\theta_{1}=15^{\circ}$ and $\theta_{2}=27^{\circ}$.}
        \label{br4_ap}
    \end{figure*}

    \begin{figure*}[]
        \centering
        \includegraphics[width=1.\textwidth]{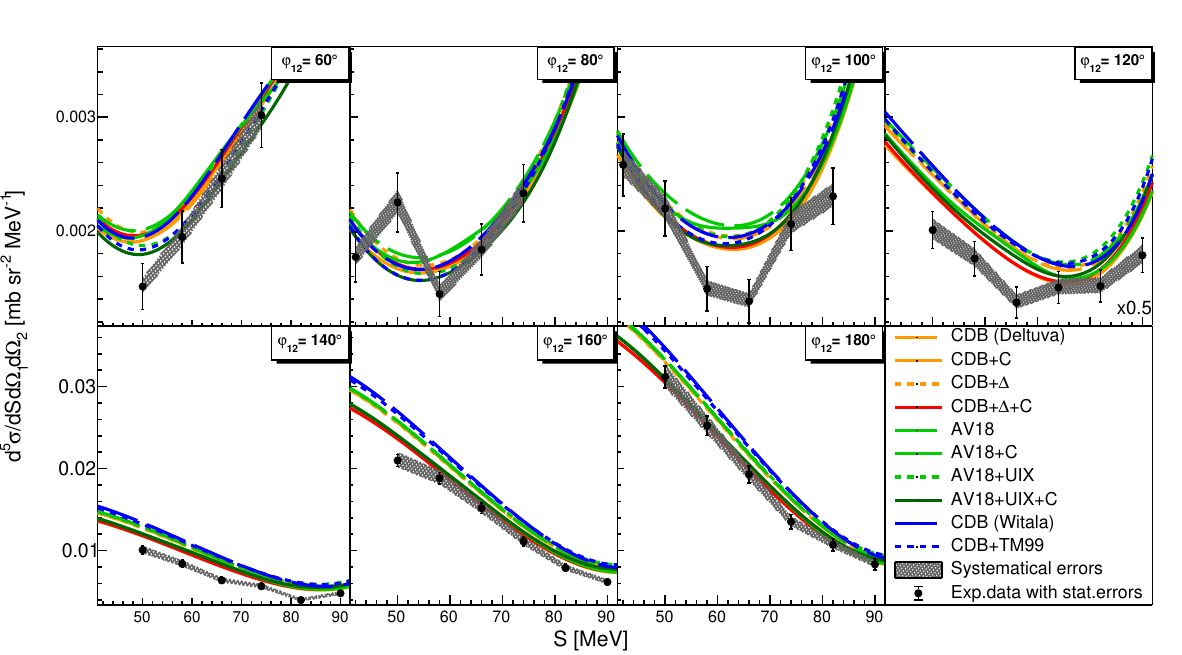}
        \caption{The same as Fig.~\ref{br1_ap} but for polar angles $\theta_{1}=15^{\circ}$ and $\theta_{2}=31^{\circ}$.}
        \label{br5_ap}
    \end{figure*}

    \begin{figure*}[]
        \centering
        \includegraphics[width=1.\textwidth]{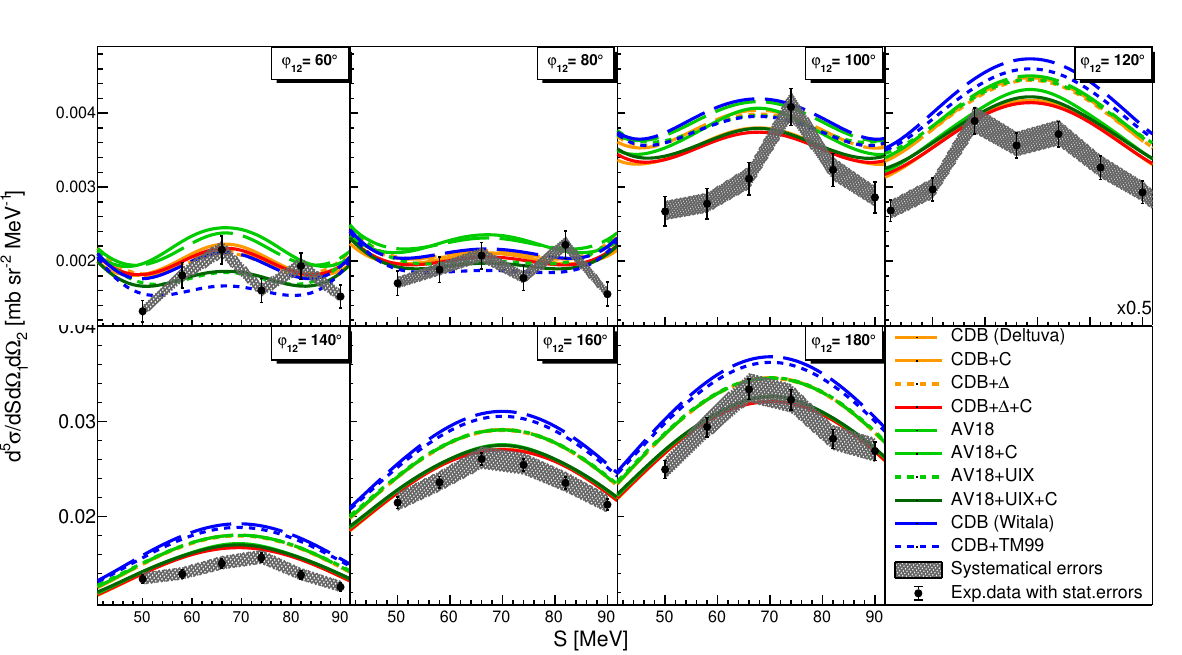}
        \caption{The same as Fig.~\ref{br1_ap} but for polar angles $\theta_{1}=19^{\circ}$ and $\theta_{2}=19^{\circ}$.}
        \label{br6_ap}
    \end{figure*}

    \begin{figure*}[]
        \centering
        \includegraphics[width=1.\textwidth]{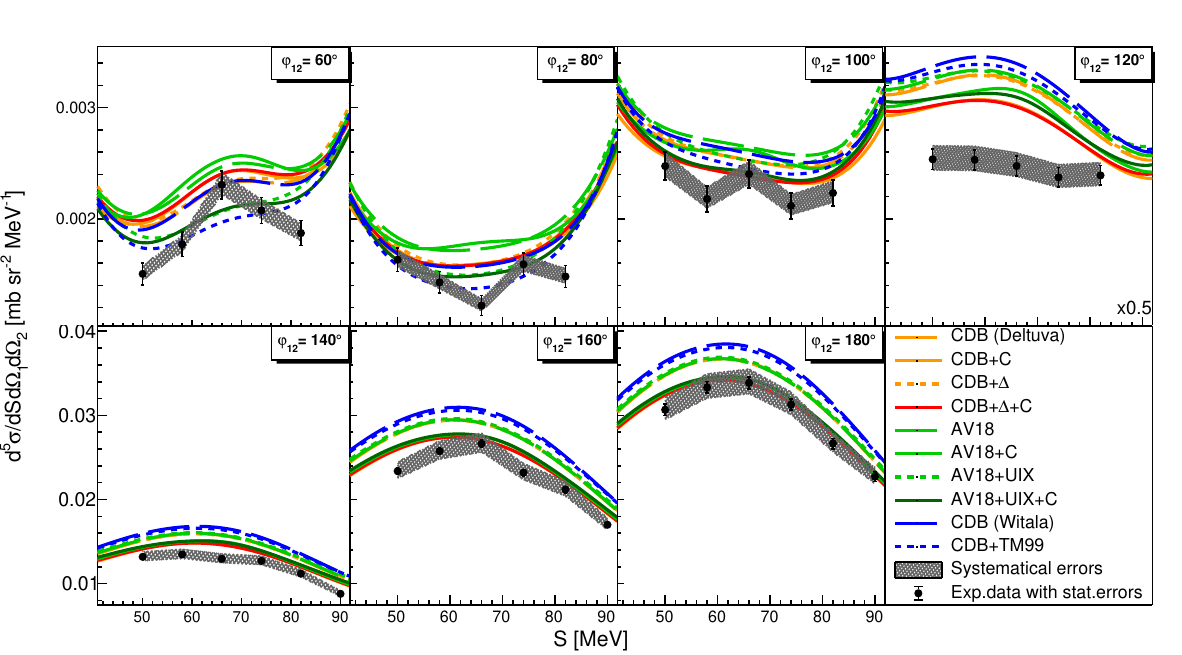}
        \caption{The same as Fig.~\ref{br1_ap} but for polar angles $\theta_{1}=19^{\circ}$ and $\theta_{2}=23^{\circ}$.}
        \label{br7_ap}
    \end{figure*}

    \begin{figure*}[]
        \centering
        \includegraphics[width=1.\textwidth]{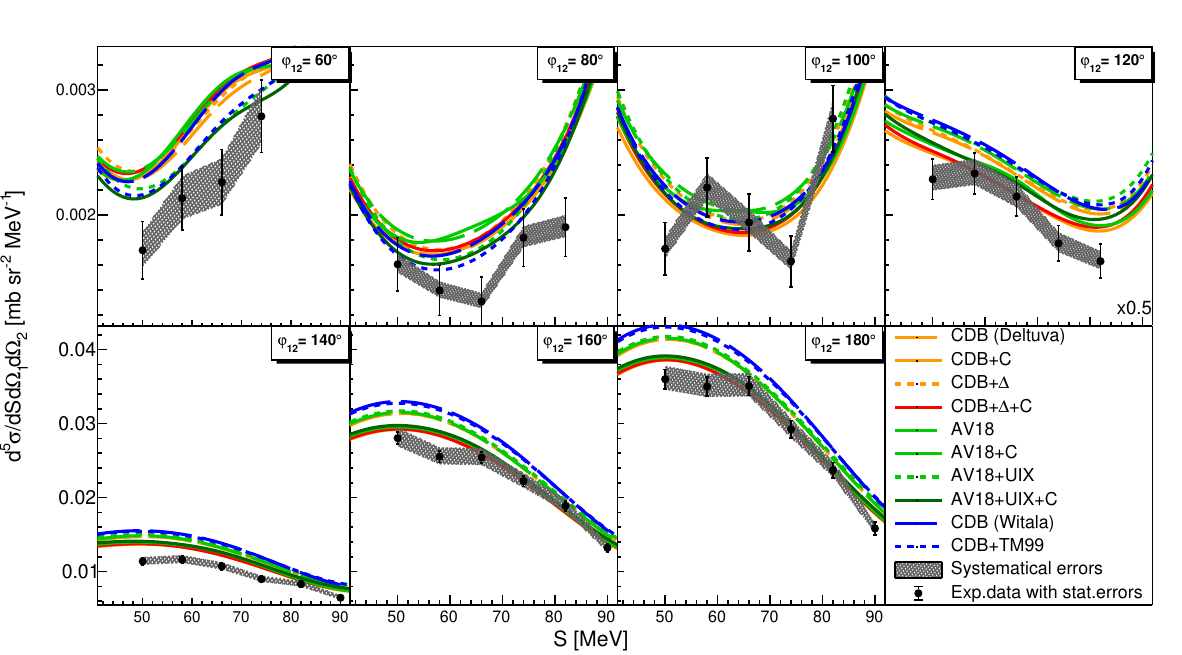}
        \caption{The same as Fig.~\ref{br1_ap} but for polar angles $\theta_{1}=19^{\circ}$ and $\theta_{2}=27^{\circ}$.}
        \label{br8_ap}
    \end{figure*}

    \begin{figure*}[]
        \centering
        \includegraphics[width=1.\textwidth]{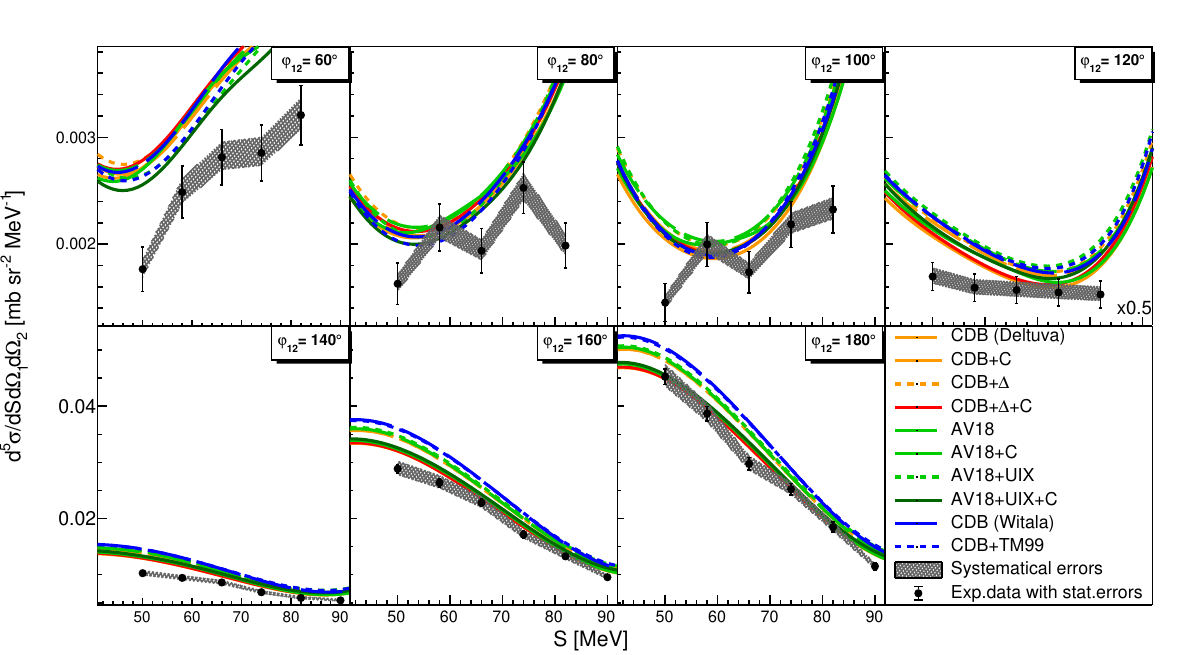}
        \caption{The same as Fig.~\ref{br1_ap} but for polar angles $\theta_{1}=19^{\circ}$ and $\theta_{2}=31^{\circ}$.}
        \label{br9_ap}
    \end{figure*}

    \begin{figure*}[]
        \centering
        \includegraphics[width=1.\textwidth]{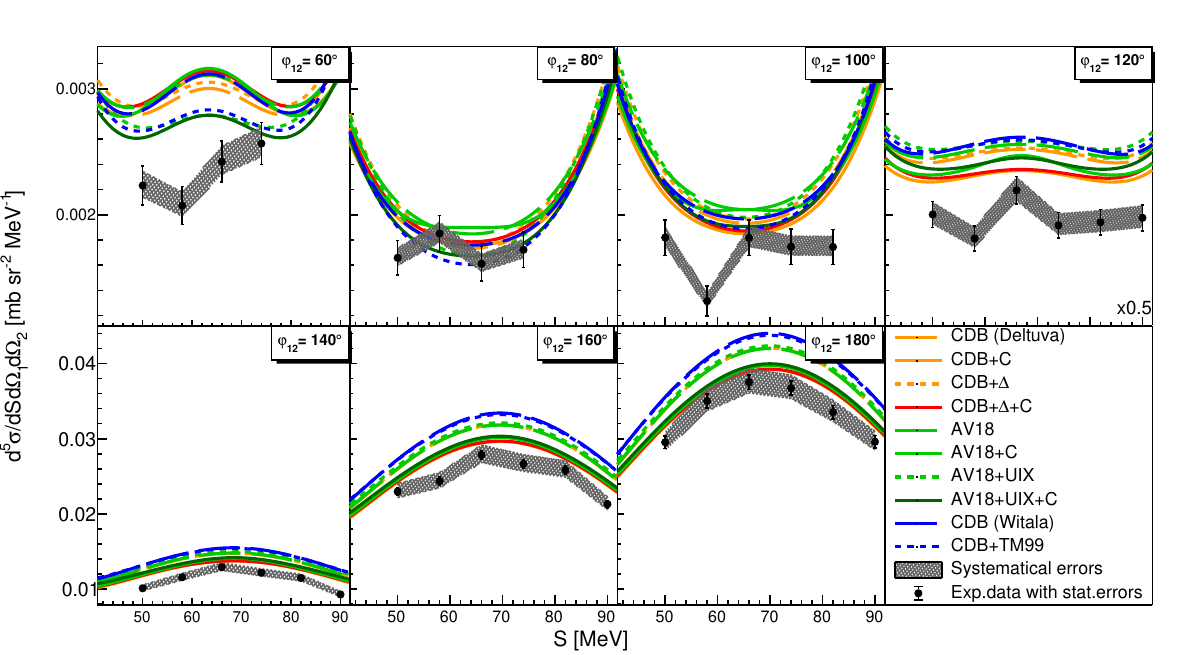}
        \caption{The same as Fig.~\ref{br1_ap} but for polar angles $\theta_{1}=23^{\circ}$ and $\theta_{2}=23^{\circ}$.}
        \label{br10_ap}
    \end{figure*}

    \begin{figure*}[]
        \centering
        \includegraphics[width=1.\textwidth]{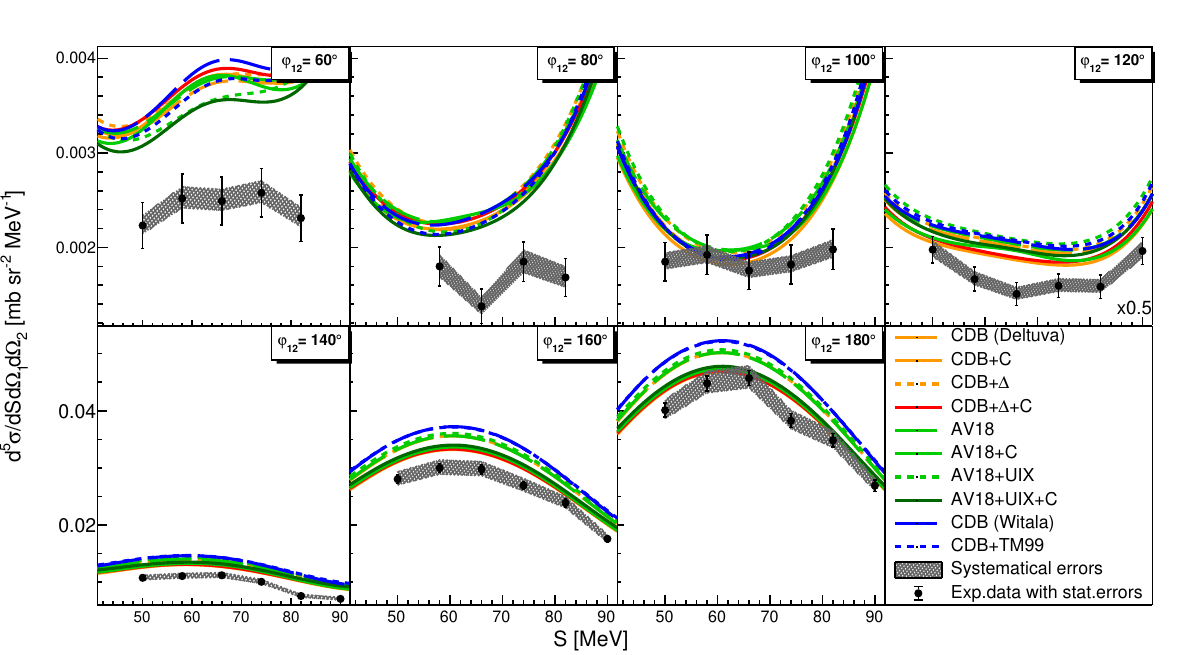}
        \caption{The same as Fig.~\ref{br1_ap} but for polar angles $\theta_{1}=23^{\circ}$ and $\theta_{2}=27^{\circ}$.}
        \label{br11_ap}
    \end{figure*}

    \begin{figure*}[]
        \centering
        \includegraphics[width=1.\textwidth]{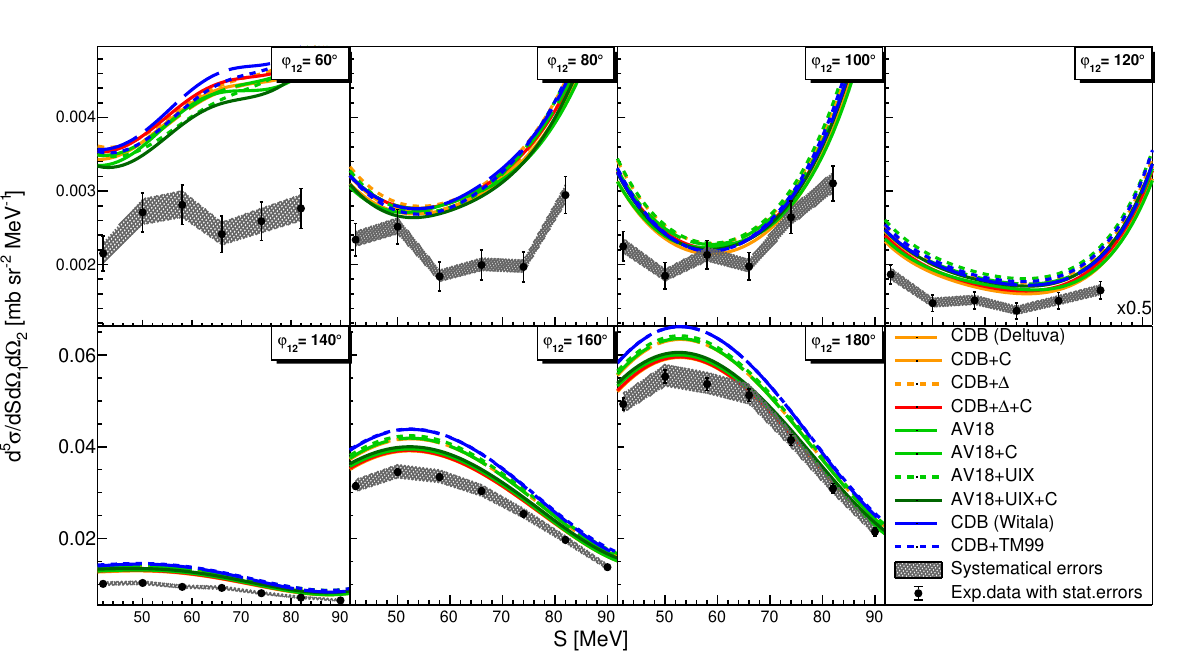}
        \caption{The same as Fig.~\ref{br1_ap} but for polar angles $\theta_{1}=23^{\circ}$ and $\theta_{2}=31^{\circ}$.}
        \label{br12_ap}
    \end{figure*}

\nocite{*}

\bibliography{bibliography}

\end{document}